\documentstyle[12pt, epsf]{article}
\def\rs { ${\bf R}^{10}\times S^1 $}

\begin{document}
\begin{titlepage}
\rightline{Imperial/TP/96-97/38}
\def\today{\ifcase\month\or
        January\or February\or March\or April\or May\or June\or
        July\or August\or September\or October\or November\or December\fi,
  \number\year}
\rightline{hep-th/9703118}
\vskip 1cm
\centerline{\Large \bf  BPS bound states, supermembranes,  }
\centerline{\Large \bf and T-duality in M-theory \footnote{
Lectures given at the APCTP
Winter School ``Dualities of Gauge and String Theories",
Korea, February 1997.}
}
\vskip 0.8cm
\centerline{\sc Jorge G. Russo }

\centerline {{\it Theoretical Physics Group, Blackett Laboratory}}
\centerline{{\it Imperial College, London SW7 2BZ, U.K. }} 
\centerline{{jrusso@ic.ac.uk }} 

\centerline{\sc Abstract}
This is an introductory review on the eleven-dimensional description
of the BPS bound states of type II superstring theories, and on the
role of supermembranes in M-theory.
The first part describes  classical solutions of 11d supergravity
which upon dimensional reduction and T-dualities give
bound states of NS-NS and R-R $p$-branes of
type IIA and IIB string theories.
In some cases (e.g. $(q_1,q_2)$ string bound states of 
type IIB string theory), these non-perturbative objects
admit a simple eleven-dimensional description in terms of 
a fundamental 2-brane. 
The BPS excitations of such
 2-brane are calculated and shown
to exactly match
the  mass spectrum  for the BPS $(q_1,q_2)$ string bound states.
Different 11d 
representations of the same bound state
can be used to provide  inequivalent (T-dual) descriptions
of the oscillating BPS states.
This permits to test T-duality beyond perturbation theory
and, in certain cases, to evade membrane instabilities by going to
a stable T-dual representation.
We finally summarize the results indicating in what regions of the
modular parameter space a supermembrane description for M-theory on ${\bf R^9}\times {\bf T^2}$ seems to be adequate.

\end{titlepage}
\newpage

\def\beq{\begin{equation}}
\def\eeq{\end{equation}}
\def\bea{\begin{eqnarray}}
\def\eea{\end{eqnarray}}
\renewcommand{\arraystretch}{1.5}
\def\ba{\begin{array}}
\def\ea{\end{array}}
\def\bce{\begin{center}}
\def\ece{\end{center}}
\def\nn{\noindent}
\def\nonu{\nonumber}
\def\pbx{\partial_x}

\def \Q {{\cal Q}}
\def \H {\td H}
\def \la{\longrightarrow}
\def \up {\uparrow}
 \def \upa {\uparrow}
 \def \nea {\nearrow}
\def \pa {\Vert}
\def\ma{\mapsto}
\def\inv{^{-1} }
\def \K {{\tilde K}}
\def\ww {\omega _{km } }
\def\mn { {(-k,-m)} }

\def\np {{  Nucl. Phys. }}
\def \pl {{  Phys. Lett. }}
\def \mpl {{ Mod. Phys. Lett. }}
\def \prl {{  Phys. Rev. Lett. }}
\def \pr  {{ Phys. Rev. }}
\def \ap  {{ Ann. Phys. }}
\def \cmp {{ Commun.Math.Phys. }}
\def \ijmp {{ Int. J. Mod. Phys. }}
\def \jmp {{ J. Math. Phys.}}
\def \cqg {{ Class. Quant. Grav. }}


\def\ptl{\partial}
\def\del{\partial }
\def\a {\alpha}
\def\b{\beta}
\def\ga{\gamma} 
\def\Ga{\Gamma}
\def\de{\delta} \def\De{\Delta}
\def\ep{\epsilon}
\def\vep{\varepsilon}
\def\ze{\zeta}
\def\et{\eta}
\def\t{\theta} \def\Th{\Theta}
\def\vth{\vartheta}
\def\io{\iota}
\def\ka{\kappa}
\def\l{\lambda} 
\def\La{\Lambda}
\def\rh{\rho}
\def\s{\sigma} \def\Si{\Sigma}
\def\ta{\tau}
\def\up{\upsilon} 
\def\Up{\Upsilon}
\def\p{\phi} 
\def\Ph{\Phi}
\def\vph{\varphi}
\def\ch{\chi}
\def\ps{\psi} 
\def\Ps{\Psi}
\def\om{\omega} 
\def\Om{\Omega}

\def \la{\longrightarrow}
\def\lbr{\left(}
\def\rbr{\right)}
\def\half{\frac{1}{2}}
\def\ha { {\textstyle {1\over 2}} }
\def\td{\tilde }
\def\ov{\over }

\def\vol#1{{\bf #1}}
\def\nupha#1{Nucl. Phys., \vol{#1} }
 \def\CMP#1{Comm. Math. Phys., \vol{#1} }
\def\phlta#1{Phys. Lett., \vol{#1} }
\def\phyrv#1{Phys. Rev., \vol{#1} }
\def\PRL#1{Phys. Rev. Lett., \vol{#1} }
\def\prs#1{Proc. Roc. Soc., \vol{#1} }
\def\PTP#1{Prog. Theo. Phys., \vol{#1} }
\def\SJNP#1{Sov. J. Nucl. Phys., \vol{#1} }
\def\TMP#1{Theor. Math. Phys., \vol{#1} }
\def\ANNPHY#1{Annals of Phys., \vol{#1} }
\def\PNAS#1{Proc. Natl. Acad. Sci. USA, \vol{#1} }

\def\mapa#1{\smash{\mathop{\longrightarrow }\limits_{#1} }}

\section{ Introduction}

In type II superstring theories, classical solutions representing solitons 
can be classified into two types: the NS-NS and the R-R solitons,
according to whether they carry charge of gauge fields originating
from the NS-NS or R-R sector of the theory (for recent reviews and references
see \cite{duality}-\cite{TSE}).
In certain cases, these solitons can form bound states \cite{john, witten, polch} and  the corresponding
classical solution preserves some of the original supersymmetries.
These are the BPS bound states.
The bound states can be {\it marginal} (or  {\it at threshold}),
meaning that they have zero binding energies (typically, 
$M_{1+2}=M_1+M_2$), or else {\it non-marginal}
(or {\it non-threshold}), with a finite binding energy
(viz. $M_{1+2}=\sqrt{M_1^2 + M_2^2}$~). In this paper
we will focus on these last ones \cite{rutse}
(different studies of supersymmetric M-brane solutions can be found e.g.
 in refs.~ \cite{papd}-\cite{jerome}).

Type IIB superstring theory can be connected to M theory  
by the sequence
\beq
{\rm Type\ IIB\ on\ } S^1 \ \mapa{T}\ {\rm  Type \ IIA\ on\ }S^1  \ 
\mapa{\rm R_1\to \infty }\  {\rm 10d\ Type\ IIA} \mapa{g^2\to \infty }\ 
{\rm M\ theory}
\label{eq:33}
\eeq
In the presence of several isometries, 
the path from eleven dimensions to  type IIB string theory is not unique.
The same classical solution of eleven-dimensional supergravity can be
used to obtain different solutions of type IIA superstring
theory, according to the direction we reduce. Furthermore,
if there is an extra isometry,
the T-duality transformation connecting type IIA and type IIB theory can be
done in a direction which is either perpendicular or parallel to the brane.
There are many  possible paths which are not manifest
in this sequence and, as a result, there are 
inequivalent eleven-dimensional representations of the same type IIB solutions.

The equations of motion of type
IIB superstring theory are symmetric under the action of $SL(2,{\bf R})$
transformations \cite{sen, hull}.
Starting with the fundamental NS-NS string solution in type
IIB superstring theory, by an  $SL(2,{\bf R})$ transformation one can obtain
a general solution which  represents the 
 bound state of NS-NS and R-R strings. 
From the viewpoint of eleven dimensions, such transformation
simply corresponds to a reparametrization; 
the $SL(2, {\bf R})$ transformations are viewed as the modular transformations
of the target torus \cite {john,aspin}.
This illustrates a  characteristic of M-theory:
in a number of cases, 
what is non-perturbative from string theory viewpoint can be described
in eleven dimensions within the domain of perturbation theory.
Many complicated-looking solutions of string theory take a remarkably simple form when
lifted to eleven dimensions. 
This is because the 11d counterpart of performing a U-duality transformation
and adding Kaluza-Klein charges
to a NS-NS soliton does not further complicate the solution; it gives
the same solution in terms of some rotated coordinates  or with a 
momentum boost.
In particular, as we shall discuss, 
the string bound states admit an exact description
in terms of a fundamental 2-brane with a certain charge and a momentum boost. 

In the next section,
we will discuss the eleven dimensional origin of different non-marginal BPS configurations
of type II string theories  as classical solutions of the supergravity
effective field theory in eleven dimensions. 
Following \cite{rutse,trusso}, we will 
explicitly construct the solutions which upon reduction and
dualities give different $p$-brane bound states
(other interesting discussions on non-marginal bound states in type II theories can be found in refs.~\cite{myers,papa}).
In lower dimensions, these can be obtained from pure NS-NS configurations by U-duality transformations (i.e. combining $O(d,d)$ and $SL(2,{\bf R})$
transformations).

In section 3 the connection between type IIA superstrings and supermembranes wrapped on \rs\ will be investigated by using the light-cone Hamiltonian
approach. The question that will be addressed (investigated
in \cite{russo}) is whether at small compactification radius
the only light excitations are those contained in the  type IIA
superstring spectrum. It turns out that this is not the case, and there are extra
quantum states, which do not decouple in the zero radius limit.
We shall return to this point in sect. 4, where
it will be argued that standard supermembrane theory is not suitable 
in certain regimes.

Given that the string-string bound states correspond to fundamental
2-branes in eleven dimensions, it is of interest to determine the spectrum
of oscillations of these 2-branes.
This can be done by using the light-cone 
Hamiltonian approach of supermembrane theory.
Although this is a non-linear theory, it will be shown that
it is exactly solvable in a certain limit \cite{rutse}.
In the BPS sector, a remarkable correspondence with the bound state
spectrum  proposed by Schwarz \cite{john} will be found.
We will also calculate the spectrum for another (T-dual) representation
of the same string-string bound state \cite{trusso}, and find 
that in the BPS sector the  spectra corresponding to the T-dual backgrounds
match, which may be regarded as a test of T-duality 
beyond perturbation theory (since the matching  holds true
for BPS quantum states of masses $\a' M^2=O(1/g^2)$).
Some surprises arise
in the non-BPS sector: the assumption of exact T-duality in M-theory
implies that supermembrane theory can only be an 
adequate description in some corners of the modular parameters. 
Possible corrections to the membrane Hamiltonian outside
these corners can be deduced by using the T-dual representation.

\section{Eleven-dimensional origin of BPS bound states}

We will now map 10d supergravity solutions of type IIA and IIB
into 11d supergravity.
The notation will be as follows:

\noindent {\it Type IIB supergravity multiplet:}
$$
NS\otimes NS:\ \ \ \{ \phi_B, \ B^{(1)}_{\mu\nu }, \ g_{\mu\nu } \}
$$
$$
R\otimes R: \ \ \ \{ \chi ,\  B^{(2)}_{\mu\nu }, \ D_{\mu\nu\rho\s } \}
$$
\noindent {\it Type IIA supergravity multiplet:}
$$
NS\otimes NS:\ \ \ \{ \phi_A, \ B_{\mu\nu }, \ \tilde g_{\mu\nu } \}
$$
$$
R\otimes R: \ \ \ \{ A_\mu ,\  A_{\mu\nu\rho } \}
$$
\noindent {\it $d=11$ supergravity multiplet}
$$
\{ g^{(11)}_{\hat \mu \hat \nu }, \ C_{\hat \mu \hat \nu \hat \rho } \}\ 
$$
\bigskip
Type IIA and 11d supergravity solutions are connected
by dimensional reduction:
\bea
&& ds^2_{11}=g^{(11)}_{\hat\mu\hat\nu }dx^{\hat\mu } dx^{\hat\nu }=
e^{-2\phi_A /3} \tilde g_{\mu\nu }dx^\mu dx^\nu +e^{4\phi_A/3}
(dy -A_\mu dx^\mu )^2\ ,
\nonu \\
&& A_{\mu\nu\rho}=C_{\mu\nu\rho}\ ,\ \ \ \ B_{\mu\nu }=C_{\mu\nu y}\ .
\label{eq:12}
\eea
The T-duality map between type IIA/IIB solutions is given  by \cite{berg}
$$
\td g_{yy}=g_{yy}^{-1} \ ,\ \ \ \ e^{2\phi_A}={ e^{2\phi_B}\ov g_{yy}}\ ,
$$
$$
\td g_{\mu\nu }=g_{\mu\nu } - g_{yy}^{-1}\big( g_{y\mu } g_{y\nu } -
B_{y\mu }^{(1)}B_{y\nu }^{(1)}\big)\ ,\ \ \ \ 
\td g_{y \mu }= { B_{y\mu }^{(1)}\ov g_{yy}}\ \ ,
$$
$$
A_{y\mu\nu }={2\ov 3} \big[ B_{\mu\nu }^{(2)} + 2g_{yy}^{-1}
B_{y[\mu }^{(2)}g_{\nu ]y} \big]\ ,\ \ 
$$
$$
A_{\mu\nu \rho}= {8\ov 3} D_{y\mu\nu\rho}+\epsilon ^{ij}
B_{y[\mu }^{(i)}B_{\nu \rho ] }^{(j)}
+g_{yy}^{-1}
\epsilon ^{ij}B_{y[\mu }^{(i)}B_{y \nu }^{(j)} g_{\rho ]y} \ ,
$$
$$
B_{\mu \nu }= B_{\mu\nu }^{(1)} + 2 g_{yy}^{-1}
B_{y[\mu }^{(1)}g_{\nu ]y}\ ,\ \ \ \ 
B_{y\mu }={ g_{y\mu }\ov g_{yy}}\ ,
$$
\beq
A_\mu =- B_{y\mu }^{(2)} +\chi B_{y\mu }^{(1)}\ \ ,\ \ \ \ \ A_y=\chi \ .
\label{eq:3}
\eeq

Let us now consider type IIB superstring theory on ${\bf R^9}\times S^1$, 
and let
$y_1$ be the coordinate of $S^1$ with radius $R_1'$.
The basic object that  will be used as starting point is the 
fundamental string solution \cite{dgh},  which in the string frame is given by
\def\td{\tilde }
\bea
&&ds^2_{10B}=\td H_2^{-1} \big( -dt^2+dy_1^2\big) + dx_idx_i\ ,
\ \ \ \ e^{2\phi }= \td H_2^{-1} \ ,
\nonu \\
&&B_{ty_1}^{(1)}= -\td H_2^{-1}(\td H_2-1)\ ,\ \ \ \ B_{ty_1}^{(2)}=0\ ,
\nonu \\
&&\td H_2=1+{\td Q\ov r^6}\ ,\ \ \ \ r^2=x_ix_i\ .
\label{eq:fss}
\eea
[Throughout indices $i$ will run over all remaining directions.]
By a straightforward application of the   IIA/B T-duality map (\ref{eq:3}),
one finds the type IIA background that is obtained by
a T-duality transformation in $y_1$  \cite{horo,hortse}:
\bea
&&ds^2_{10A}=-dt^2+dy_1^2+(\td H_2-1)(dt+dy_1)^2+dx_idx_i\ ,\ 
\nonu \\
&&e^{2\phi_A}=1\ ,\ \ \ A_\mu=A_{\mu\nu\rho}=B_{\mu\nu}=0\ ,
\eea
representing a gravitational wave moving along $y_1$. 
In the string theory, for T-duality to be a symmetry, $y_1$ must have  periodicity $R_1=\a'/R_1'$.

Using eq. (\ref{eq:12}),  we now lift this solution to $d=11$, obtaining
\bea
&&ds^2_{11}=-dt^2+dy_1^2+dy_2^2+(\td H_2-1) (dt+dy_1)^2+dx_idx_i\ ,\ 
\nonu \\
&&C_{\mu\nu\rho }=0\ ,
\label{eq:11}
\eea
again representing a gravitational wave. Note that, since $y_1$ is periodic,
the momentum $\td Q$ must be quantized in units of $1/R_1$,
which with the correct normalization gives
$$\td Q= c_0 {n\ov R_1} \ ,\ \ \ \ c_0={\kappa_9^2\ov 3\omega_7}\ ,
\ \ \ \
\kappa_9 ^2 =2\pi R_1 \kappa_{10}^2\ ,\  \ \ \omega_7={\pi^4\ov 3}\ .
$$
From ten-dimensional viewpoint, this quantization condition arises as the Dirac quantization implied by the existence of the five brane.

The solution (\ref{eq:11}) preserves half of the supersymmetries, and it 
admits a simple generalization where
$\td H_2-1 \equiv W$ is replaced by $W=W(u,x^i)$, with $\ptl^2_x W =0$
\cite{hortse}.
A solution of interest is $W=f_i(u)x^i$, representing right-moving waves
in the direction of the boost. 
For a solution carrying momentum along a given direction, 
supersymmetry is preserved
by adding waves moving in the direction of the momentum
 \cite{hortse, dabwal}.

In \cite{john}, the $(q_1,q_2)$ string bound states were obtained by
applying an $SL(2,{\bf Z})$ transformation to the fundamental string solution. The corresponding transformation in $d=11$ turns out to be
quite simple: take the solution  (\ref{eq:11}) and make a rotation in $y_1,y_2$ plane.
This gives 
\bea
&&ds^2_{11}=-dt^2+dy_1^2+dy_2^2+(\td H_2-1) (dt-\cos\theta dy_1-\sin\theta
dy_2)^2+dx_idx_i\ ,\
\nonu \\
&&W={\td Q_q \ov r^6}\ ,\ \ \ 
\td Q_q=c_0\sqrt{ {l_1^2\ov R_1^2}+  {l_2^2\ov R_2^2} }=
\td Q\sqrt{q_1^2+q_2^2\tau_2^2 }\ ,\ \ 
\\
&& \cos\theta= {q_1\ov \sqrt{q_1^2+q_2^2\tau_2^2} }\ ,\ \ \ \  
\tau_2={R_1\ov R_2}=e^{-\phi_{B0}}\ .
\label{eq:coseno}
\eea
Here $l_1,l_2$ are integer numbers corresponding to the Kaluza-Klein momenta in $y_1,y_2$ directions, and we have introduced the
relatively prime integers $q_1,q_2$ by $(l_1,l_2)=n(q_1,q_2)$.
Upon reduction we now find
\bea
&&
ds^2_{10A}=K^{1/2}\big[ -K^{-1} d\td t^2 +d\td y_1^2
+dx_idx_i\big]\ ,\ \ \ \ 
\nonu \\
&&e^{4\phi /3}=K\ ,\ \ \ \ K=1+\sin^2\theta \ W \ ,
\nonu \\
&&\td t={1\ov \sin\theta} \big( t-\cos\theta y_1\big)\ ,\ \ \ \ 
\td y_1={1\ov \sin\theta} \big( y_1-\cos\theta t\big)\ . 
\eea
This represents a 0-brane with a {\it finite} boost with velocity
$v=\cos\theta $ \cite{rutse}.
By T-duality along $y_1$, we find the corresponding type IIB solution
\bea
&&
ds^2_{10B}=K^{1/2}\big[ \td H_2^{-1} \big(-dt^2+dy_1^2\big) + dx_idx_i\big]\ ,
\nonu \\
&& e^{2\phi }= \td H_2^{-1} K^2\ ,\ \ \ \ \chi =\sin\theta\cos\theta WK^{-1}\ ,
\nonu \\
&& 
B_{ty_1}^{(1)}= -\cos\theta W \td H_2^{-1}\ ,\ \ \ \ 
B_{ty_1}^{(2)}= -\sin\theta W \td H_2^{-1}\ .
\label{eq:uno}
\eea
For $\cos\theta=0 $, one has $K=1$ and the solution reduces
to the fundamental string solution (or $1_{NS}$); for
$\cos\theta=1 $, one has $K=\td H_2 $ and the solution represents the R-R string (or $1_R$).
Schematically, we have obtained the sequence
\beq
\nea  \ \mapa{\rm red}\ 0_R \upa \ \mapa{T_{y_1} }\ 1_R + 1_{NS}
\label{eq:flec}
\eeq

Another basic object in eleven dimensions is the 2-brane given
by  \cite{duste}
\bea
&& ds^2_{11}=H_2^{-2/3}\big[ -dt^2+dy_1^2+dy_2^2\big] +H_2^{1/3} dx_idx_i\ ,
\nonu \\
&&
C_3=H_2^{-1} dt\wedge dy_1\wedge dy_2\ ,\ \ \ \ H_2=1+{Q\ov r^6}\ ,
\eea
where $Q=c_0{w_0R_1\ov\a' }$, and $w_0$ represents the winding number of the 2-brane
around the target torus $y_1,y_2$.
Upon reduction, this gives
\bea
&& 
ds^2_{10A}=H_2^{-1}\big[-dt^2+dy_1^2\big] +dx_idx_i \ ,
\nonu \\
&&
e^{2\phi }=H_2^{-1}\ ,\ \ \ B_{y_1 t}=C_{y_2y_1 t}= H_2 ^{-1}\ ,
\eea
i.e., the fundamental string solution in type IIA superstring theory.
After $T_{y_1}$-duality, we get a plane wave in type IIB.
Summarizing:
\beq
2\ \mapa{\rm red}\  1_{NS}  \ \mapa{T_{y_1} }\ \upa
\eeq
Combining with the previous result  (\ref{eq:flec}), we now 
start with the $2\nea $ background given by \cite{rutse}
\bea
&&ds^2_{11}=H_2^{-2/3}\big[ -dt^2+dy_1^2+dy_2^2+(\td H_2-1) (dt- dz_1)^2 \big] + H_2^{1/3} dx_idx_i\ ,\
\nonu \\
&& C_3=H_2^{-1} dt\wedge dy_1\wedge dy_2\ ,\ \ \ \ 
z_1=y_1 \cos\theta +y_2\sin\theta\ ,
\label{eq:repA}
\eea
and get
\beq
2\nea  \ \mapa{\rm red}\ 1_{NS}+ 0_R + \upa \ \mapa{T_{y_1} }\ 
\upa + 1_R+ 1_{NS}
\label{eq:todo}
\eeq
The final solution in type IIB superstring theory 
represents the 1/4 supersymmetric $(q_1,q_2)$ string bound state with a boost
along $y_1$ direction, with the metric
\beq
ds^2_{10B}=K^{1/2}\big[ \td H_2^{-1} \big(-dt^2+dy_1^2
+(H_2-1)(dt-dy_1)^2
\big) + dx_idx_i\big]\ ,
\eeq
and all other fields as in eq.  (\ref{eq:uno}).

The fundamental string background 
 (\ref{eq:fss}) solves the equations of motion at $r\neq 0$. In order
to solve the equations of motion at $r=0$, a string-like source term
needs to be added \cite{dgh}, with tension
$$
T= {\omega_7 \ov \kappa^2_{10} }\td Q\ .
$$
For the $(q_1,q_2)$ string background, the relevant string source must have
tension
$$
T_q={\omega_7 \ov \kappa^2_{10} } \td Q_q=T\sqrt{q_1^2+q_2^2\tau_2^2}\ .\ 
$$
This has led Schwarz to propose that, at weak coupling, the spectrum of oscillations
of the $(q_1,q_2)$ string bound state must be given 
by the usual free string spectrum, with $T\to T_q $.
The mass formula is then
\bea
M^2_{\rm IIB} &&= {w_0^2 \ov {R'_1}^2 }+ 4\pi^2T^2_q n^2 {R'_1}^2  
+4\pi T_q(N_R+N_L) 
\nonu \\
&&={l_1^2{R_1'}^2\ov {\a ' }^2 } + {w_0^2\ov {R_1'}^2} +
{l_2^2\ov R_2^2 } + {4\pi T \ov n} 
\sqrt{ l_1^2 + l_2^2 e^{-2\phi_0 } }(N_R+N_L)\ ,
\label{eq:bosti}
\eea
$$
N_R- N_L=w_0 n\ . \ 
$$
The zero mode part of this formula (which already
contains information about the non-trivial
tension) is in exact correspondence with the mass formula of
a wrapped supermembrane \cite{john} . 
To see this, we first perform
a { T-duality} transformation and get
$$
M^2_{\rm IIA}= {l_1^2\ov R_1^2} + {l_2^2\ov R_2^2 }+
{ w_0^2 {R_1}^2\ov {\a ' }^2 } + {4\pi T \ov n} 
\sqrt{ l_1^2 + l_2^2 e^{-2\phi_0 } } (N_R+N_L)\ .
$$
For a wrapped membrane, we have
$$
M^2= ( w_0AT_3)^2 + {l_1^2\ov R_1^2} + {l_2^2\ov R_2^2 }+ ...\ ,
$$
where $w_0,~A,~T_3$ are the membrane winding number, area and tension respectively,
and dots represent the oscillator contributions that we are for the moment ignoring. By writting
$$
w_0AT_3= 4\pi^2 w_0 R_1R_2 T_3= 2\pi R_1 w_0 T\ ,\ \ \ \ T\equiv 2\pi R_2 T_3\ ,
$$
and identifying $T$ with the string tension,
we see that the zero mode part of the membrane mass formula agrees 
with the string bound state counterpart.

The mass formula  (\ref{eq:bosti}) should be exact for BPS states.
In particular, it should not receive additional corrections as the type IIA
string coupling $g^2=R_2^2/\a' $ is varied from $0$ (where eq.  (\ref{eq:bosti}) applies) to $\infty $. 
In the BPS sector, it is meaningful to compare
the spectrum  (\ref{eq:bosti}) with the BPS excitations of the 2-brane.  
For generic BPS states, $N_R=w_0 n\neq 0$; the oscillator
contribution is indeed non-vanishing and we must  
 calculate it in order to establish the equivalence 
between mass formulas \cite{rutse}. This will be done in  section 4.

Starting with the $(q_1,q_2)$ string bound state with an extra isometry in the coordinate $y_3$,
we can apply $T_{y_3}$ duality and then lift the resulting
solution to eleven dimensions,
obtaining an alternative 11d representation of the type IIB
string-string bound state. Schematically
$$
1_{NS}+1_R+\upa \ \mapa{T_{y_3}}\ 1_{NS}+2_R+\upa  \ \mapa{\rm lift}\ 
2\upa  
$$
One obtains \cite{trusso}
\bea
&& ds^2_{11}=\td H_2^{-2/3}\big[ -dt^2+dy_1^2+dz_2^2
+(H_2-1) (dt-dy_1)^2
\big] +\td H_2^{1/3} (dz_3^2+dx_idx_i)\ ,
\nonu \\
&&
C_3=\td H_2^{-1} dt\wedge dy_1\wedge dz_2\ ,\ \  
\label{eq:repB}
\\
&&z_2=y_2\cos\theta+y_3\sin\theta\ ,\ \  
\ \ z_3=-y_2\sin\theta+y_3\cos\theta\ .
\nonu
\eea
This is a 2-brane with a momentum boost $w_0/R_1'$ along $y_1$, with
one leg on $y_1$ and the other wrapped around the $(q_1,q_2)$ cycle
of the torus $(y_2,y_3)$.
When $w_0=0$ (corresponding to the type IIB string bound state with
zero momentum), this becomes a {\it static } background, whereas
in the representation (A) the $w_0=0$ case has non-zero momentum. 
There exists no reparametrization
which connects both backgrounds;  
from the viewpoint of supergravity effective field theory
the two solutions are inequivalent. 
They are, however, of the same form, where the roles of $Q$ and $\td Q $
(i.e. winding number and total momentum) are interchanged, but in addition
the membrane  is wrapped in a different way around the 3-torus.
The question that remains is whether, in M-theory, these two
inequivalent geometries can be physically equivalent, just as  in string
theory  $\s $-model backgrounds related by T-duality represent the same conformal field theory (see sect. 4).

Before considering more complicated examples, it is useful to summarize the  rules governing the basic duality
operations:

\noindent (1)  T-duality along the boost: 
plane wave  $ \leftrightarrow $  FS.
T-duality in transverse direction: FS and plane wave unchanged.

\noindent (2) Parallel T-duality:  D$p$-brane $\la $ D$(p-1)$ brane.  
Transverse T-duality: D$p$-brane $\la $    D$(p+1)$-brane.

\noindent (3) Reduction along  boost direction $\la $ 
 R-R charge. Reduction along direction orthogonal to boost $\la $ 
boosted brane.

\noindent (4)  S-duality $\phi_B \to -\phi_B $: R-R and NS-NS 1,5 branes are exchanged.
(The 3-brane remains invariant). Boost unchanged. 
In $d=11$, S-duality corresponds to the simple reparametrization
of  exchanging  the 11d direction with a T-duality direction.

The $2+0$ bound states of type IIA superstring theory can
also be connected to the $SL(2,{\bf Z})$ string bound state, 
provided there are two extra isometries. The required operations
are 
$$
(1_{NS}+1_{R})_2\ \mapa{T_\bot ^2} \ 1_{NS} + 3_R \ \mapa{S} \  1_R +3_R \ 
\mapa{T_{||}}
\ 0_R+2_R\ \mapa{\rm lift}\ 2 \ma 
$$
The final 11d configuration $2 \ma $ indicates a transversely boosted 2-brane.
The bound state $1_R+3_R$ background that one obtains in this
way is given by \cite{rutse}
\bea
&& ds^2_{10 B} =  \tilde H^{1/2}_2 [ \tilde H_2\inv ( -dt^2 + dy_1^2) + 
    K\inv (dy_2^2 + dy_3^2) +  dx_i dx_i ]  \ , \ 
\nonu \\
 &&e^{2\p }=   \td H_2    K\inv \  \ ,\ \ \ \  
D_{ty_1y_2y_3} = \sin \t\ (\td H_2-1) K\inv \ , \ 
\\  
&&B^{(2)}_{t y_1 }  =  - \cos \t\ (\td H_2 -1)  \td H_2 ^{-1}  \ , \ \  
  B^{(1)}_{y_2 y_3 }  =  \sin \t  \cos \t\  (\td H_2-1) K^{-1}  \ .   
\nonu
\eea
T-duality converts it into the $(0_R+2_R)_1$ bound state represented by
\bea
&&ds^2_{10 A} =   \td H_2^{1/2} [ - \td H_2\inv dt^2 + 
K\inv (dy_2^2 + dy_3^2)   + dy_1^2  +  dx_s dx_s ]  \ , 
\nonu \\ 
&&e^{2\p }=   \td H_2^{3/2}  K\inv   \ , \ \  \ \ \ 
C_{ty_2y_3} = - \sin \t \ (\td H_2-1) K\inv \ , 
\label{eq:21} \\
&&A_{t }  =  - \cos \t\ (\td H_2 -1) \td H_2\inv \ , \ \  \ \ \ 
  B_{y_2 y_3 }  =  \sin \t\  \cos \t\   (\td H_2-1) K^{-1}  \ .   
\nonu
\eea
This interpolates between the 0-brane ($q_2=0, \ K=1$) and the 2-brane
wrapped around $y_2,y_3$ ($q_1=0, \ \td H_2 =K$).
The coordinate $y_1$ does not play any special role, and in the final geometry
the restriction of the extra isometry can be removed, i.e. $y_1$ can be added
to the $x_i$, obtaining the $2+0$ background which is spherically symmetric
in all 7 transverse coordinates.

Lifting the type IIA solution (\ref{eq:21}) to  $d=11$ one obtains
\bea
&&ds^2_{11} =  K^{1/3} \big[  K\inv  (-  d\td t^2 + 
dy_2^2 + dy_3^2) 
  +  d\td y^2_{11}  +  dx_i dx_i \big]  \ , 
\nonu \\
&&dC_3 =   dK\inv \wedge   d\td t\wedge dy_2 \wedge dy_3  \ ,   \ \
\eea
where 
$$ \td t\equiv {1\ov \sin \t} ( t - \cos \t\ y_{11}) \ , \ \ \ \ 
\td y_{11} \equiv  {1\ov \sin \t} ( y_{11} - \cos \t\  t ) \ .
$$
This is a 2-brane  
 boosted  to a subluminal  velocity  $v= \cos \t \leq  1 $ 
 in the isometric  transverse  direction  $y_{11}$  (or $2 \ma $). 

The generalization to the case $(1_{NS}+1_{R}+\upa )_2$, i.e. when
the original string configuration has non-vanishing momentum, is straightforward, and gives the sequence
$$
(1_{NS}+1_{R}+\upa )_2\ \mapa{T_\bot S T_{||} }\  
2_R\bot 1+0_R\ \mapa{\rm lift }\ 2 \bot 2 \ma 
$$
Explicit formulas for the backgrounds are given in \cite {rutse}.
Note that, since we have previously connected the $(q_1,q_2)$ string bound state
to a fundamental 2-brane, in the space with two extra isometries $2 \bot 2 \ma $ is  dual to a single 11d 2-brane .


Analogous backgrounds can be derived by including 5-branes.
For example, the $5_{NS}+5_R$ non-marginal bound state 
can
be obtained by an $SL(2,{\bf Z}) $ rotation on the solitonic
$5_{NS}$ brane of type IIB theory. 
Another background of interest is the 1/8 supersymmetric
bound state $5_{R}+1_{R}+\upa $, which has been useful 
for the construction of $D=5$ extremal black holes with regular horizons
\cite{SV,TT,CM}.
To obtain this, one starts with the 1/8 supersymmetric 11d background $5\bot 2+\upa $ \cite{tsey,kleb} and
perform the duality operations:
$$
5\bot 2+\upa\ \mapa{\rm red}\ (5+1)_{NS}+\upa \ \mapa{T_{ y_1}}\
5_{NS}+\upa +1_{NS}\ \mapa{S}\ (5+1)_R +\upa
$$
For the D-brane  bound state $(1+5)_R+\upa $ ($n=3,4,5,6$):
$$
ds^2_{10B}= \big( H_1H_5\big) ^{-1/2} \big[-dt^2+ dy_1^2 +
(H_2-1)\big( dt-dy_1 \big) ^2 \big] +\big( {H_1\ov H_5}\big)^{1/2} dy_ndy_n
$$
$$
+  \big( H_1H_5\big) ^{1/2} dx_idx_i\ ,\ \ \ \ \ e^{2\phi }={H_1\ov H_5} \ ,
$$
$$
H_{1,5}=1+{Q_{1,5}\ov r^2}\ ,\ \ \ \ \ H_2=1+{Q\ov r^2}\ .
$$
 Reduction to $d=5$ gives
$$
ds^2_5 = -\lambda ^{-2/3} (r) dt^2 + \lambda ^{1/3} (r) 
(dr^2+r^2 d\Omega ^2_3)\ ,
$$
$$
 \lambda  (r) = H_1H_5H_2\ .
$$
The Bekenstein-Hawking entropy of this black hole can then be compared
with the statistical entropy derived by counting D-brane microstates
in the weak coupling limit \cite{SV,CM}. 

Generalizing  the previous construction involving the 2-brane, we now start with $5\bot 2+\nea $,
to obtain a $d=10$ configuration with 5 charges (4 of which are independent), representing the bound state
$$
(1+5)_{NS} + (1+5)_R +\upa
$$ 
It includes the special cases:
$ 1_{NS} + 1_{R}+ \upa $\ , \ \ 
$(1+5)_R+\upa $\ , \  \ $(1+5)_{NS}+\upa $ \ ,\ \   $5_R+5_{NS}+\upa $,
and  it thus provides a unified description of various 1-brane and 5-brane bound state configurations of type IIB theory.
The corresponding solution is given by
\bea
d s^2_{11} = && H^{2/3}_5 H^{1/3}_2
 \big(  H_5\inv H_2\inv   [-  dt^2  
 +  dz^2_1    + 
  (H_1-1) (dt -  dz_1)^2 ]  
\nonu \\
 &&   + \  H_5\inv  dy_n dy_n +    H_2\inv  dz^2_2       
+  dx_i dx_i \big) \ ,    
\\
dC_3 =  && dH_2\inv\wedge  dt \wedge dy_1 \wedge dy_2  
 + *dH_5  \wedge dz_2 \  ,  
\eea
$$
z_1=y_1\cos\theta + y_2 \sin\theta\ ,\ \ \ \ 
z_2=-y_1\sin\theta + y_2 \cos\theta\ .
$$
Reduction down to $d=5$ gives a family of regular extremal black holes
related to the simplest NS-NS and R-R ones by U-duality.

\vfill\eject

\section{Type IIA strings from membranes on \rs }

The bosonic part of the supermembrane action is given by \cite{bergsh}
$$
S={T_3\ov 2}\int d^3\s \bigg[ \sqrt{-\gamma } \gamma ^{ij}
\del _i X^{\hat \mu} \del X^{\hat \nu} g_{\hat \mu\hat \nu }\ - \ \ha
\sqrt{-\gamma } +\ {1\ov 6} \epsilon ^{ijk}\del_i  X^{\hat \mu}\del_j X^{\hat \nu}
\del_k X^{\hat \rho} C_{\hat \mu \hat \nu \hat \rho }(X)\bigg]\ ,\ \ \ 
$$
$$
\s_i = (\s , \rho, \tau )\ ,\  \ \ \hat \mu =0,1,...,10\ .
$$
The
 connection with the superstring theory is
through the so-called double-dimensional reduction ansatz \cite{duf}.
For flat $\s $-model couplings, this is the
essentially the statement that  type IIA superstring theory is recovered
provided:

\noindent 1) Make partial gauge choice $\ \ X^{10}=\rho R_{0}$ ;

\noindent 2) Assume $\del _\rho X^\mu =0 $\ , \ \ \ $\mu=0,1,...,9 $\ .

\noindent Then one gets the action
$$
S={T\ov 2} \int d^2\s \big[  \sqrt{-h} h^{\a\b } \del_\a X^\mu \del_\b X_\mu
+ \epsilon ^{\a\b }  \del_\a X^\mu \del_\b X^\nu B_{\mu\nu }(X) \big]
\ ,\ 
$$
$$
 T\equiv 2\pi T_3 R_{0}\ .
$$
In order to check  whether 
superstring theory indeed arises as a small radius limit of 
supermembrane theory on \rs , one should be able to show that 
all quantum states containing oscillation modes in $\rho $ 
become heavy and decouple as  $R_{0}\to 0$, leaving
only the type IIA superstring spectrum as the light excitations
of the theory. 
Nevertheless, it will be seen 
that in this limit there remain extra quantum states,
which are associated with flat directions of the membrane quartic potential.
For a toroidal membrane wrapped on ${\bf R}^9\times {\bf T}^2$,
these extra quantum states will be removed in the limit
that one torus cycle is shrunk to zero, keeping 
the type IIB string coupling finite and small (see sect. 4).

The problem can be investigated by using 
the  light-cone Hamiltonian \cite{bst,dewitt}. Let us first
consider membranes on ${\bf R}^{11}$.
Defining as light-cone coordinates
$$
X^\pm= {X^0\pm X^{10} \ov \sqrt {2} }\ , 
$$
the bosonic part of the Hamiltonian
takes the form (fermions can be easily incorporated, see e.g.
\cite{dewitt,russo})
\bea
&& H =  2\pi ^2 \int d  \sigma d\rho \left [
 P _ a   ^ 2 +
{ T_3^2\over 2 }
( \{ X ^ a, X ^ b \} ) ^ 2\right ] \ ,
\\
&& \{ X, Y \} = \partial _ {\s } X  
\partial _ {\rho } Y - \partial _ {\rho } X  
\partial _ {\s } Y \ ,\ \ \ \ a=1,2,...,9\ .
\nonu
\eea
\def\n{ {\bf n} }
\def\m{ {\bf m} }
This Hamiltonian has a residual gauge symmetry under the group
of area-preserving diffeomorphisms associated with the membrane topology
under consideration.
These are reparametrizations of the form
$$
\s^\a \la \s^{\prime\a }=\s^\a +\xi^\a (\s,\rho )\ ,\ \ \ \ \xi^\a =\epsilon^{\a\b } \del_\b \phi \ ,\ \ \ \ \s^\a =(\s,\rho) \ ,
$$
$$
X^a (\s ')=X^a (\s )+ \xi^\a \del_\a X^a= X^a (\s )+\{ X^a,\phi \}\ .
$$
If $\{ Y_{\n }(\s ,\rho )\} $ denotes a complete set of functions on the
membrane surface, 
the Lie bracket $\{ Y_{ \n }, Y_{ \n '} \}$ describes the  
area-preserving diffeomorphisms algebra associated with the
topology, and the $Y_\n$ (or the corresponding quantum operator $L_\n $, see below) generate arbitrary transformations,
e.g. $\delta_\n X^a=\{ X^a,Y_\n \}$.
Here we will be concerned with toroidal membranes, for which
\beq
 Y_{\n }(\s ,\rho )= e^{ik\s +im\rho }\  , \ \ \   \n =(k,m)\ ,\ \  
\s , \rho \in [0, 2\pi )\ .
\eeq
[Throughout, indices $m,n $ are used for  Fourier modes in $\rho $, whereas
$k,l$ are associated with Fourier modes in $\s $.]
The corresponding Lie bracket gives 
\beq
\{ Y_{ \n }, Y_{ \n '} \} = f_{\n \n ' \m } Y^\m =
- (\n \times \n ' ) Y_ {\n +\n '} \ ,
\label{eq:lieb}
\eeq
$$
\ \ \ \ \n \times \n ' = k m' -m k' \ .
$$
The truncated or regularized version, 
where $k,m=0,...,N-1$, $(k,m)\neq (0,0)$, generates the algebra of 
$SU(N)$ \cite{fairl}.
The same regularized Hamiltonian \cite{dewitt} arises in 
Yang-Mills quantum mechanics (dimensional reduction of 
 super YM from $D=9+1$
to $D=0+1$~), and also in describing the short distance dynamics of $N$ D0-branes. In a sense, the connection with   D-branes 
has clarified the appearance of Yang-Mills quantum mechanics \cite{witten},
which in the original derivation of the matrix model \cite{dewitt} 
was somewhat mysterious. A number of new aspects have also been ellucidated
 in the recent formulation of ref.~\cite {banks} (see also \cite{dvv}).

Let us now consider membranes moving on \rs .
We introduce the dimensional parameter 
$$
\a'= (4\pi^2 R_0 T_3)^{-1}\ .
$$
 For a configuration that wraps once along $S^1$,
we have  
\beq
X^{10}(\s ,\rho +2\pi )=X^{10}(\s ,\rho )+ 2\pi R_0\ ,
\eeq
so that
\beq
X^{10}=R_0 \rho + \td X^{10} \ ,
\eeq
where $\td X^{10} $ is single-valued. The single-valued part can be removed
by the light-cone gauge choice 
$X^+= {X^0+ \td X^{10} \ov \sqrt {2} }=x^+ + \a'  p^+\tau $.

The Hamiltonian takes the form \cite{russo}
($X'(\s )\equiv\del _\s X(\s )$ )
$$
H=H_0+ H_{\rm int} \ ,
$$
where
\bea
&& H_0=2\pi ^2 \int d  \sigma d\rho \left [
  P _ a   ^ 2 +
T_3^2 R_0^2 (X_a')^2 \right ] \ ,
\\
&& H_{\rm int }=\pi ^2 T_3^2 \int d  \sigma d\rho 
\big( \{ X ^ a, X ^ b \} \big) ^ 2 \ .
\eea
If we now expand in Fourier modes in $\rho $ 
$$
X^a(\s , \rho ,\tau )= \sum _m X_m^a(\s ,\tau  ) e^{im \rho } \ ,
\ \ \ P^a (\s ,\rho ,\tau )= {1\ov 2\pi }
\sum _m P_m^a(\s ,\tau ) e^{im \rho } \ ,
$$
$$
[X^a_m(\s ), P^b_n(\s ' )]=i\delta _{m+n }\delta ^{ab}\delta (\s -\s ' )\ 
,\ \ \ \ X^\dagger_m= X_{-m}\ ,
\ \ \ P^\dagger_m= P_{-m}\ ,
$$
the Hamiltonian becomes
\beq
H_0=\pi T^2 \int d\s 
\sum _m \left [
 T^{-2} P^a_{m} P^a_{-m} + X_{m}^{a\prime }{X_{-m}^{a\prime }} \right ]
\ ,\ \ \ \ T\equiv 2\pi R_0T_3\ , 
\label{eq:aster}
\eeq
\beq
H_{\rm int}= {\pi T^2\ov R_0^2} \int d\s 
\sum _{m,n,p} \big[
mp ({X_n^{a\prime }} X_p^a) ( X^{b\prime }_{-m-n-p} X_m^b) -
np (X_n^a X_p^a) ( X^{b\prime }_{-m-n-p} X_m^{b\prime })
\big]\ .
\label{eq:asterr}
\eeq
The symmetry of area-preserving diffeomorphisms can be gauge fixed by setting
\beq
X^{9}=X^{9}_K(\rho )\ ,\ \ \ \ 
X^{9}_K(\rho )=\sum _{m} X_{(0,m)} ^{9} e^{im\rho }\ 
\label{eq:cari}
\eeq
(the $Y_{(0,m)}=e^{im\rho }$ generate a Cartan subspace).
In addition, there is a local constraint
\beq
\del_\a X^-={1\ov p_0^+}\del_\a X^a \dot X^a\ .
\label{eq:18}
\eeq
By taking the curl, one gets the condition
\beq 
\{ X^a, \dot X^a\} =\del_\s X^a \del _\rho \dot X^a -
\del_\rho X^a \del _\s \dot X ^a
\equiv 0\ ,
\label{eq:19}
\eeq
or, in phase-space variables,
\beq
\{ X^a, P^a\}\equiv 0\ .
\label{eq:20}
\eeq
The Fourier components
\beq
L_\n ={1\ov 4\pi ^2} \int d\s d\rho\ Y^\n \{ X^a, P^a\}\ 
\eeq
are generators of the algebra of area-preserving 
diffeomorphisms (\ref{eq:lieb}).
By a gauge transformation, we can always rotate one of the coordinates,
say $X^{9}$, into the Cartan subspace, as in eq.~(\ref{eq:cari}).
Then eq.~(\ref{eq:20}) takes the form
\beq
L_{\bf m} ^{\rm T}=-\sum_{\n '} \sum _{\n \in K} 
X^{9}_\n P^{9}_{\n '} f_{\n \n ' {\bf m} }\ ,\ \ 
\label{eq:22}
\eeq
where
\beq 
L_{\bf m} ^{\rm T}\equiv 
{1\ov 4\pi ^2} \int d\s d\rho\ Y^{\bf m} \{ X^i, P^i\} \ .
\eeq
We note that the $P_{\n '}^{9} $, $\n ' \in K$, are absent from this formula.
The constraint (\ref{eq:22})   determines  
$\td P^{9}(\s ,\rho ) \equiv \sum _{\n \not \in K} P_\n^{9} Y_\n (\s ,\rho )$
in terms of the $X^i , \ P^i $ and $X_K^{9}$.
By formally inverting eq. (\ref{eq:22}), one gets
\beq
\td P^{9}_{\n }=\{ F^{-1}\} _{ {\bf m} \n } L_{\bf m} ^{\rm T}\ , \ \ \ \ 
F _{\n{\bf m }}= - \sum _{\n ' \in K} X^{9}_{\n'} f_{\n ' \n  {\bf m} }\ .
\eeq
The determinant of $F$ vanishes when some of the eigenvalues of $X^{9}_{\n'} $
coincide, i.e. 
at the boundary of the Weyl chamber \cite{dewit}.
In the present case of a membrane wrapped on \rs , the relation is
always invertible in the large radius limit \cite{russo}.
 At small radius,
the discussion of instability modes
is nevertheless not affected, since one can always choose suitable
wave packets with support in 
the interior of the Weyl chamber.

Since $X^-$ is single-valued, eq. (\ref{eq:18}) also implies the global constraints
\bea
&&{\bf P^{(\s )} }={1\ov 2\pi \a' } 
\int _0^{2\pi } d\s \ \del_\s X^a \dot X^a \equiv 0\ ,
\label{eq:26}\\
&&{\bf P^{(\rho )} }= {1\ov 2\pi \a' }
\int _0^{2\pi } d\rho \ \del_\rho X^a \dot X^a \equiv 0\ .
\label{eq:27}
\eea
The operators $ {\bf P^{(\s )} },\ {\bf P^{(\rho )} } $ generate
translations in $\s $ and $\rho $, respectively.
By virtue of eq.~(\ref{eq:19}) , the integrals in (\ref{eq:26}) and  
(\ref{eq:27}) are independent of the contours.
In particular, one readily checks that
\beq
\del_\rho {\bf P^{(\s )} }=0\ ,\ \ \ \ \ \ \ \del_\s {\bf P^{(\rho )} }=0\ .
\label{eq:28}
\eeq
By making use of the properties (\ref{eq:28}), we can write
${\bf P^{(\s )} } ,\ {\bf P^{(\rho )} }$ in the more convenient form:
\bea
&&{\bf P ^{(\s )} } 
={1\ov 4\pi ^2\a'  }\int _0^{2\pi } d\rho \int _0^{2\pi } d\s \  \del_\s X^a \dot X^a \ ,\ 
\\
&&{\bf P^{(\rho )} }={1\ov 4\pi ^2\a'  }\int _0^{2\pi } d\s \int _0^{2\pi }
d\rho \ \del_\rho X^a \dot X^a \ .\ 
\eea
These equations will later be used to write
${\bf P^{(\s )} }$ and ${\bf P^{(\rho )} }$ in terms of mode operators.

The Hamiltonian (\ref{eq:aster}), (\ref{eq:asterr}) is naturally organized
as an infinite sum of free string theory Hamiltonians labelled by $m$.
The interaction grows with $m$; strings with $m\neq 0$ are
the analogue of Kaluza-Klein modes, which  decouple from   low-energy physics
at small compactification radius.
In this Hamiltonian approach, the double-dimensional reduction  procedure  corresponds
to dropping all modes $X_m^a(\s )$ with $m\neq 0$, 
and setting  the Kaluza-Klein momentum
$p^{10}=\int d\rho P^{10}(\rho)$ to zero \cite{russo}.
What remains is 
$$
\a'  H_{\rm red }={T\ov 2 }\int d\s 
 \left [ T^{-2}P^i_{0} P^i_{0} +
{X_{0}^i}'{X_{0}^i}' \right ]  \ ,\ \ \ \ \a' \equiv (2\pi T)^{-1}\ .
$$
which is nothing but the  string  theory Hamiltonian.

World-volume time translations are generated by 
$\tilde H=\a'  ( H_0 + H_1)$. 
Regarding $H_1$ as a perturbation, the equations of motion of the
unperturbed Hamiltonian $H_0$ give
$$
\del_\s^2 X_m^a= \del_\tau ^2 X_m^a\ .
$$
The solution satisfying the periodicity condition $X^a(\s +2\pi )=X^a(\s )$
is given by
\beq
X_m^a(\s ,\tau )= x_m^a +{\textstyle{\a' } } p_m^a\tau
+ {\textstyle     {i   \sqrt{ {\a' \ov 2} }     }        }
\sum_{k\neq 0 }   {1\ov k }\bigg(
\a_{(k,m)}^a e^{-i k(\tau -\s )} + 
\tilde \a_{(k,m)}^a e^{-i k(\tau +\s )} \bigg) \ ,
\eeq
\beq
[\a_{(k,m)}^i, \a_{(l,n)}^j]=k \delta _{k+l} \delta _{m+n}\delta ^{ij} \ ,
\ \ \ \ \a_{(k,m)}^9=\td \a_{(k,m)}^9=0\ ,
\eeq
$$
\a_{(k,m)}^\dagger=\a_{(-k,-m)}\ ,\ \ \ x_m^\dagger=x_{-m}\ ,
\ \ \ p_m^\dagger= p_{-m} \ .
$$


The Hamiltonian 
has potential valleys along $x_m^a$, corresponding to the
constant modes in the coordinate $\s $.
Indeed, the $x_m^a$ do not appear in $H_0$ nor in $H_{\rm int}$.
The extra term $H_{\rm int}$ in the potential contains flat directions
along  all Cartan directions \cite{dewit};  the $Y_{(0,m)}=e^{im \rho}$
(associated with the $x_m^a$) span a Cartan subspace.
The $x_m^a $ with $m\neq 0$ are  the only directions that are not stabilized
by the winding contributions, and 
they are  responsible of
the instabilities of the supermembrane on \rs \cite{russo}.
One can  construct wave packets in these directions which
move off to infinity, and this
holds true for any value of the radius $R_0$.

Let us now incorporate $H_{\rm int}$. By expanding 
$X_m^a (\s ),\ P_m^a(\s )$ in terms of mode operators
$$
 X_m^a (\s )=\sqrt{\a'  } \sum _{k} X_{(k,m)}^a e^{i k \s }\ ,\ \ \ 
 P_m^a(\s )={1\ov 2\pi \sqrt{\a' } } \sum _{k\neq 0} 
P^a_{(k,m)} e^{ ik \s }\ ,\ \ \ 
$$
we obtain
$$
\a'  H_0=\ha \sum _{\n } \big[  P^a_{\n }  P^a_{-\n }
+k^2 X^a_{\n }  X^a_{-\n } \big]\ ,
$$
$$
\a'  H_{\rm int}= {1\ov 4g^2}\sum (\n_1 \times \n_2) (\n_3\times \n_4)
X_{\n_1}^a X_{\n_2}^bX_{\n_3}^aX_{\n_4}^b\ ,
$$
$$
g^2\equiv {R_0^2\ov \a' }=4\pi^2R_0^3T_3\   ,
$$
where the sum runs over $\n_1, \n_2, \n_3$, and $\n_4=-\n_1-\n_2-\n_3$.
The parameter
$g^2$ is exactly the type IIA string coupling that is obtained
upon reduction of 11d supergravity. Let us now consider
the properties of the system as the string coupling $g^2$
is changed at fixed $\a '$.
The  mass  operator is given by
\beq
M^2=2p^+_0p^-_0-(p_0^{a})^2=2 H_0+2 H_{\rm int}-(p_0^{a})^2\ .
\eeq
$H_{\rm int}$ is positive definite, and any state $|\Psi \rangle $ with 
$\langle \Psi | H_{\rm int} |\Psi \rangle \neq 0$ will have infinite
mass in the zero coupling limit.
The only states that survive in the zero coupling limit
(with $T_3\to \infty $ so that 
$T =2\pi R_0 T_3$ remains fixed) are
those containing excitations in a Cartan subspace of the area-preserving
diffeomorphism algebra, so that $H_{\rm int}$ drops out from
$\langle \Psi | M^2 |\Psi \rangle $
(to be precise, from $X_{(k,m)}^a, P_{(k,m)}^a$ one  introduces 
creation and annihilation operators
$a_{(k,m)}^a, a_{(k,m)}^{a\dagger }$ in the standard way; any state
made with a set  $\{ a_{(k,m)}^{a\dagger } \}$ whose associated 
$\{ Y_{(k,m)}\} $ are non-commuting  has infinite 
mass in the zero coupling limit).

The $X_{(k,0)}^i$ generate a Cartan subspace, implying that
the full type IIA superstring spectrum survives,
$$
\langle \Psi _{\rm IIA}| H_{\rm int} |\Psi _{\rm IIA}\rangle = 0\ .
$$ 
In addition,  there are excitations
in other  directions which also remain.
This includes wave packets made with the zero models $x_m^a, p_m^a$
(that is, states made of  $X_{(0,m)}^a, \ P_{(0,m)}^a$ oscillators, whose 
corresponding generators $Y_{(0,m)}$ also span a Cartan subspace).

 Let us now consider the infinite coupling limit, $R_0 \to \infty $, $T_3\to 0$, with
$T=2\pi R_0 T_3$ fixed. In this limit, $g^2\to \infty $,
so  the  term $ H_{\rm int}$ can be dropped, and the Hamiltonian
becomes that of an  infinite set of harmonic oscillators labelled by
$(k,m)$. 
Note that this is true only for a membrane that
wraps around the compact dimension;  in this limit 
the quartic terms 
in the potential are negligible in relation to the quadratic terms.
For a membrane that does not wrap around $S^1$, there is no quadratic term,
and at any radius the dynamics is governed by the quartic terms.

The (bosonic part of the) mass spectrum takes the form
$$
\ha \a'   M^2= \ha\a'  \sum _{m\neq 0} p_m^a p_{-m}^a
+ \sum_{m=-\infty }^\infty \sum_{k=1} ^\infty
\big[  \alpha _{(-k,-m)}^i  \alpha _{(k,m)}^i
+\tilde \alpha _{(-k,-m)}^i  \tilde \alpha _{(k,m)}^i \big] \ 
$$
\beq
= \ha\a'  \sum _{m\neq 0} p_m^a p_{-m}^a
+  N^+_\s +    N^-_\s \ .
\label{eq:50} 
\eeq
In the $R_0\to \infty $ limit, the fact that the 
standard membrane spectrum is continuous is simply  understood:
the center-of-mass momenta of the strings with $m\neq 0$,
$p_m^i=\int d\s P_m^i(\s )$ take continuous values, since
they are governed by the free
particle Hamiltonian 
$H_{\rm free}=\ha p_m^i p_{-m}^i =\ha \big( 
p_m^{(1)}\big) ^2+\ha \big( p_m^{(2)}\big) ^2$,
where $p_{\pm m}=p^{(1)}_m\pm i p^{(1)}_m$.

We recall that we have gauge fixed the symmetry of
area-preserving diffeomorphisms by setting $\a^{9}_{(k,m)}=0$. 
 The physical Hilbert space is spanned by states made of the 
transverse excitations
$\alpha _{(-k,m)} ^i, \ \td \alpha _{(-k,m)} ^i $ (and the fermion 
partners $\ S _{(-k,m)} ^r,\ \td S _{(-k,m)} ^i$), 
with $k>0$ and  $m\in {\bf Z}$.
In terms of mode operators, the  global constraints 
$ {\bf P^{(\s )} }={\bf P^{(\rho )} }=0 $  become the level matching
conditions:
\bea
&& N^+_\s={N^-_\s}\ ,\ \ 
\\
&&  N^+_\rho -{  N^-_\rho}=l_0\ ,
\label{eq:l0}
\eea
where  $l_0\in {\bf Z}$ is 
the Kaluza-Klein charge, $p_0^{10}=l_0/R_0\ $, which in the ten-dimensional
theory is viewed as a R-R charge.
Explicit expressions for $ N^\pm _\rho $ will be given below 
for a more general membrane configuration.

Unlike the spectrum of superstring theory,  
the membrane spectrum 
contains the infinite tower of Kaluza-Klein quantum states 
carrying Ramond-Ramond charges. The ten-dimensional mass operator
is given by 
$$
M^2_{10D}=M^2+ (p_0^{10})^2\ .\ \ \ \ 
$$
 Kaluza-Klein states with $l_0\neq 0$
will thus have $M_{\rm 10D}=O(1/R_0)$, or $\sqrt{\a' } M_{\rm 10D}= O(1/ g )$.
 Note that these R-R quantum states with $l_0\neq 0$ 
 must contain oscillations with $m\neq 0$ in order to
satisfy the  constraint (\ref{eq:l0}). 
There are, in addition, other 
quantum states with vanishing R-R charge $l_0=0$,
having nonetheless masses  of order $1/g$. These are states containing excitations of 
$a^\dagger _{(k,m)}$ with $m\neq 0$; the  
 $O(1/g)$ mass originates from contributions of $H_{\rm int}$.

In the regularized $SU(N)$ theory, eq.~(\ref{eq:50}) takes the form
\beq
\ha \a'   M^2 = \ha\a'  \sum _{m=1}^{N-1} p_m^a p_{-m}^a
+  N^+_\s +    N^-_\s \ . 
\eeq
It may be worth emphasizing that the continuity of the spectrum is
not associated with the center-of-mass momentum $p_0^2$, which is in $M^2$. 
The $p_m^a$ operators, with $m=1,...,N-1$,
are genuine degrees of freedom of the membrane Hamiltonian, 
associated with rigid ($\s $-independent) motion in direction 
transverse to $\s $ (i.e. along $S^1$).
However, 
in the particular sector $l_0=N$ \cite{banks}, 
the $p_m^a$ (including the center-of-mass $p_0^a$) 
may be naturally interpreted as the momenta of a (bound state) system of $N$ D0-branes, whose short-distance dynamics is indeed governed by the $SU(N)$ Hamiltonian. For a  general membrane 
configuration with
non-zero winding around the target torus (see sect. 4),
the corresponding type IIA system  not only contains D0-branes
but, as we have seen,   represents a bound state of D0-branes,
a fundamental string, and a boost (i.e. $1_{NS}+0_R+\upa $,
see eq.~(\ref{eq:todo})~).

\section{Mass spectrum of 11d 2-branes }

\def\n{ {\bf n} }

In section 2 we have seen that the $(q_1,q_2)$ string bound states
admit two representations in eleven dimensions  
in terms of a fundamental 2-brane, given by eqs. (\ref{eq:repA})
and (\ref{eq:repB}) (henceforth (A) and (B)~). In this section we will calculate
the spectrum of excitations of such 2-branes. We shall start with the representation
(A), describing to a 2-brane with momentum $l_1/R_1$ and $l_2/R_2$
along the cycles of the target torus, and having winding number $w_0$
(which in type IIA string theory on ${\bf R^9}\times S^1$
becomes the winding number of a string along the circle $S^1$).
The physical spectrum of wrapped membranes of toroidal topology
has been previously investigated in \cite{bergsh,dufi} in the semiclassical
approximation. The Hamiltonian approach used here permits to have
a better control of non-linearities of the theory.

Let  $X^{1}$, $X^{2}$ be the compact coordinates with
periods $2\pi R_{1} $, $2\pi R_{2}$, and thus consider a membrane
configuration of toroidal topology wrapped in the following way:
$$
X^{1}(\s +2\pi ,\rho) = X^{1}(\s ,\rho ) + 2\pi w_0 R_{1} \ ,
$$
$$
 X^{2}(\s ,\rho +2\pi ) = X^{2}(\s ,\rho ) + 2\pi  R_{2} \ . 
$$
Hence 
$$
X^{1}(\s,\rho )=w_{0}R_{1}\s +\tilde X^{1}(\s ,\rho) \ ,\ \  
$$
$$
X^{2}(\s,\rho )= R_{2}\rho +\tilde X^{2} (\s,\rho )\ ,
$$
where $\tilde X^{1}(\s ,\rho)$, $\tilde X^{2} (\s,\rho )$ are single-valued
functions, which can be expanded in a complete set of functions
on the torus,
$$
\tilde X^{2} (\s,\rho )=\sqrt{\a'  }\sum _{k,m} X_{(k,m)}^{2} e^{ik\s +im\rho }
\ ,\ \ \ \ 
\tilde X^{1}(\s, \rho )=  \sqrt{\a'  } \sum _{k,m} X_{(k,m)}^{1} 
e^{ik\s +im\rho }\ ,
$$
$$
\a' \equiv ( 4\pi ^2 R_{2} T_3 )^{-1}\ . \ \ \ \ 
$$
The winding number that counts how many times
the membrane is wound around the target torus is given by
$$
w_0={1\ov 4\pi^2 R_{1} R_{2} }\int d\s d\rho \ \{ X^{1}, X^{2}\}  \ .
$$
A membrane with $w_0\neq 0$ is topologically protected against
usual supermembrane instabilities. We will see this explicitly in the Hamiltonian formulation; because of contributions due to winding,
flat directions in the Hamiltonian will be removed and 
the mass spectrum will  be discrete.

Let also expand the transverse fields in terms of mode operators,
\beq
X^i (\s , \rho )=\sqrt{\a'  } \sum _{k,m} X_{(k,m)}^i e^{i k \s+im\rho  }\ ,\ \ \ 
 P^i(\s, \rho )={1\ov (2\pi)^2 \sqrt{\a' } } \sum _{k,m}  
P^i_{(k,m)} e^{ ik \s + im\rho}\ .
\eeq
Separating the 
winding contributions and inserting the expansions,
the  Hamiltonian  takes the form 
$H=H_0+H_{\rm int}$ , with (~$\n=(k,m)$, $a,b=2,...,10$)
$$
\a'  H_0= \ha \a 'T_3^2 A ^2 w_0^2+ 
 {1\ov 2} \sum _\n \big[ P_\n^a P^a_{-\n}
+\ww ^2 X^a_\n X^a_{-\n }\big]
$$
$$
\a'  H_{\rm int}= {1\ov 4g^2}\sum (\n_1 \times \n_2)(\n_3\times \n_4)
  X_{\n_1}^a  X_{\n_2 }^b  X_{\n_3}^a   X_{\n_4}^b  
$$
$$
+{i\ov g}\sum mk^2 X_{2(0,m)}X^i_{(k,n)}X^i_{(-k,-n-m)}\ ,
$$
where 
$$
g^2\equiv {R_{2}^2\ov \a' }=4\pi^2R_2^3T_3\ ,\ \ \ \ \ \ 
\ww =\sqrt { k^2 + w_{0}^2 m^2 \tau_2^2 }\ ,\ \ \ \ \ \ 
\tau _2= {R_{1}\ov R_{2} }\ .
$$

Let us first investigate the connection with type IIA superstring
theory in the limit $g^2\to 0$ at fixed $\tau_2={R_1\ov R_2}$
(small torus area, $R_1R_2\to 0$).
The case discussed in sect. 3 --a membrane on \rs -- 
involved  certain subtle  points, because
the spectrum was continuum. 
Having now a discrete spectrum,   we would like to pose again 
the question
of what quantum states survive in the zero coupling limit.
The analysis of section 3 can be repeated.
From the form of the  mass operator,
$M^2=2 H_0 +2H_{\rm int}-(p_0^{a})^2$,
we see that any state $|\Psi \rangle $ with 
$\langle \Psi | H_{\rm int} |\Psi \rangle \neq 0$ will have 
$\langle \Psi | M^2 |\Psi \rangle \to \infty $ as $g^2\to 0$.
The  quantum states  with 
$\langle \Psi | H_{\rm int} |\Psi \rangle = 0$
are
again those containing excitations in a Cartan subspace of the 
area-preserving diffeomorphism algebra, 
so that $H_{\rm int}$ gives no contribution.
This includes the  type IIA superstring spectrum (in the sector
with winding number $w_0$),  and
quantum states made with $X_{(0,m)}^i, P_{(0,m)}^i$, etc.
There is, however, an important difference with respect to 
the membrane on \rs .
To make contact with perturbative type IIB string theory upon T-duality,
the {\it type IIB} string coupling $e^{\phi_{B0}}=R_2/R_1$ must also be
small. Consequently, due to the large
 term $w_0^2m^2 e^{-2 \phi_{B0}} $ in the frequency $\omega_{km}$,
 all states  containing oscillators $X_{(k,m)}^i, P_{(k,m)}^i$,
with $m\neq 0$ get a non-perturbative
mass of order $e^{-\phi_{B0}}={R_1\ov R_2}\gg 1$,
so they decouple.
Just the type IIA superstring spectrum, made with 
$X_{(k,0)}^i, P_{(k,0)}^i$,  survives.
Thus,  type IIA superstring theory is exactly  
recovered from wrapped supermembranes on $T^2$ in the limit that
$R_2\to 0$ with  fixed $e^{\phi_{B0}}=R_2/R_1 \ll 1$.
Only type IIA quantum states with $w_0=0$ are  missing
in the membrane description (some of these  states --those
with a non-vanishing momentum along a torus cycle-- will be recovered
from the T-dual membrane (B); see sect. 4.2).

In the opposite limit $g^2\to \infty $, at fixed $\a ' $ and $\tau_2$, 
 non-linear terms drop out and the system reduces to an infinite
 set of harmonic oscillators. We  now determine the mass operator
in this limit.
It is convenient to introduce  mode operators 
as follows:
\def\www {w _{ {(k,m)} } }
\beq
X^a_{(k,m)}={i\ov \sqrt {2} \www }\big[\a^a_{(k,m)}+\td \a^a_{(-k,-m)}\big]
\ ,\ \ \ \ P^a_{(k,m)}={1\ov \sqrt {2} }\big[\a^a_{(k,m)}-\td \a^a_{(-k,-m)}\big]\ ,
\eeq
$$
\big( X_{(k,m)}^a\big) ^\dagger =X_{(-k,-m)}^a\ ,\ \ \ \ 
\big( P_{(k,m)}^a\big) ^\dagger =P_{(-k,-m)}^a\ ,\ \ \ \ 
\www \equiv  \epsilon (k ) \ \ww  \ ,
$$
where $\epsilon (k )$ is the sign function.
The canonical commutation relations imply
$$
\big[ X^a_{(k,m)} , P^b_{(k',m')} \big]= i\delta_{k+k'}\delta_{m+m'}
\delta^{ab}\ ,
$$
\beq
[ \a _{ {(k,m)} }^a , \a^b_{(k',m')}]= \www \delta _{k+k'}\delta _{m+m'}\delta^{ab}
\ ,\  
\eeq
  and similar relations 
 for the $\tilde \a _{ {(k,m)} } ^a$.
In terms of these modes, the solution is given by
$$X^a (\s, \rho, \tau )= x^a +\a'  p^a \tau +
 i \sqrt{\textstyle {\a' \ov 2}} \sum_{\n \neq (0,0)}
 w_\n \inv \big[ \a _\n^a e^{ik\s +im\rho }  
+ \td \a _\n ^a e^{-ik\s -im\rho }\big] \ e^{i w_\n \tau } \ .
$$

Let the momenta in the directions $X^{1}$ and $X^{2}$
be 
$$
p_{1}={l_1\ov R_{1} }\ ,\ \ \ \ p_{2}={l_2\ov R_{2} }\ ,\ \ \ \ 
\ l_1, l_2 \in {\bf Z} \ . 
$$
\def\n{ {\bf n} }
The nine-dimensional mass operator is given by
$$
M^2= 2H-p_i^2= M_0^2 + {2\ov \a' } H_{(A)} \ ,\ \ \ \ 
M^2_0={l_1^2\ov R_1^2} + {l_2^2\ov R_2^2} + {w_0^2 R_1^2\ov 
\a  ^{\prime 2}}\ ,
$$
$$
H_{\rm (A)}= \ha \sum _\n \big( \a^a_{-\n} \a^a_{\n} + \td \a^a_{-\n} \td \a^a_{\n}\big)\ ,\ \ \ \ \ \n\equiv (k,m)\ ,\ \   a=2,...,10\ .
$$
\def\www {\omega _\n }
As in sect. 3, the level-matching conditions are determined from the
global constraints  $ {\bf P^{(\s )} }={\bf P^{(\rho )} }=0 $.
We now obtain 
$$
N_\s^+ -N_\s^-= w_0 l_1\ ,\ \ \ \ \ N_\rho^+- N_\rho ^-=  l_2 \ ,
$$
where
$$
N^+_\s = \sum _{m=-\infty }^\infty \sum _{k=1}^\infty {k\ov \ww }
\a^i_\mn \a^i_{(k,m)} \ ,
$$
$$
N^-_\s = \sum _{m=-\infty }^\infty \sum _{k=1}^\infty {k\ov \ww }
\td \a^i_\mn \td \a^i_{(k,m)} \ ,
$$
$$
N^+_\rho=\sum _{m=1}^\infty \sum _{k=0}^\infty {m\ov \ww }
\big[ \a^a_\mn \a^a_{(k,m)} + \td \a^a_{(-k,m)} \td \a^a_{(k,-m)} \big]\ ,
$$
$$
N^-_\rho=\sum _{m=1}^\infty \sum _{k=0}^\infty {m\ov \ww }
\big[ \a^a_ {(-k,m)}\a^a_{(k,-m)} + \td \a^a_{(-k,-m)} 
\td \a^a_{(k,m)} \big]\ .
$$

\subsection{BPS oscillations of bound states}

We would like to compare the membrane mass operator
that we have just obtained
\bea
&&M^2={l_1^2\ov R_1^2} + {l_2^2\ov R_2^2} + {w_0^2 R_1^2\ov 
\a  ^{\prime 2}}+ {2\ov \a' } H_{(A)}\ ,
\\
&&
N_\s^+ -N_\s^-= w_0 l_1\ ,\ \ \ \ \ N_\rho^+- N_\rho ^-=  l_2 \ ,
\eea
with the mass spectrum of the $(q_1,q_2)$ string bound state \cite{john}
\bea
&&M^2_{\rm IIB}={l_1^2\ov R_1^2} + {l_2^2\ov R_2^2} + {w_0^2 R_1^2\ov 
\a  ^{\prime 2}}+4\pi T_{(q_1,q_2)}(N_L+N_R)\ ,
\nonu
\\
&& N_R-N_L=nw_0 \ .
\label{eq:35}
\eea
Let us first consider the simplest NS-NS string, where $l_2=0$, $l_1=n$
($q_1=1$, $q_2=0$).
Then
\beq
M^2_{\rm IIB}={l_1^2\ov R_1^2} +  {w_0^2 R_1^2\ov 
\a  ^{\prime 2}}+ {2\ov\a' } (N_L+N_R)\ ,\ \ \ N_R-N_L=l_1w_0 \ .
\eeq
For BPS states, $N_L=0$, so that
\beq
M^2_{\rm IIB}={l_1^2\ov R_1^2} +  {w_0^2 R_1^2\ov 
\a  ^{\prime 2}}+ {2\ov\a' } l_1 w_0= 
\bigg({l_1\ov R_1} + {w_0R_1\ov \a' }\bigg)^2\ .
\eeq
We now identify for which states in the membrane spectrum
the corresponding background preserves some (1/2 or 1/4)
of the supersymmetries.
As mentioned in sect. 2, the only way to add waves
to the 2-brane background by preserving supersymmetry is 
along the momentum direction. 
The BPS condition can be summarized in two rules:

\noindent a) Oscillations along momentum direction.

\noindent b) Only right-moving oscillations.

\noindent In this $l_2=0$ case, since the momentum direction
is along $\s $, the first condition implies that
the relevant states are made with the $\a_{(k,0)}^i, \ \td\a_{(k,0)}^i$
(or $\a_{(k,m)}^i=\td \a_{(k,m)}^i= 0$ if $m\neq 0$).
The second condition sets $\td\a_{(k,0)}^i=0$.
It follows that
\beq
H_{\rm (A)}=N_\s^+=w_0l_1\ ,
\eeq
so that
\beq
M^2\bigg|_{\rm BPS} =\bigg({l_1\ov R_1} + {w_0R_1\ov \a' }\bigg)^2
=M^2_{\rm IIB} \bigg|_{\rm BPS}\ .
\eeq
It is interesting to note that both spectra match even before
imposing the condition (b).

Let us now consider the general case with both NS-NS and R-R charges,
$l_1, l_2\neq 0$.
Performing  the  rotation    
$$
y_1= \cos\theta \ X_{1} + \sin \theta\  X_{2} \ ,\ \ \ \  
y_2= -\sin\theta \ X_{1} + \cos \theta \ X_{2}\ ,  
$$
where $\theta $ was defined in eq.~(\ref{eq:coseno}),
we may align the momentum  with  the direction $y_1$.
The map between the target-space  torus and the toroidal 
membrane surface  is given by the zero mode part
$$
y_1^0= w_{0} R_{1} \cos \theta \ \s  +  R_{2} \sin \theta \ 
\rho\   , 
\   \ \ \ 
y_2^0= -w_{0} R_{1} \sin \theta \ \s  +  R_{2} \cos \theta
 \ \rho \ .  
$$
Consider an  oscillation mode $\a^i_{(k,m)}e^{i (k\s +m \rho )}$
$$
k\s + m\rho = \bigg( {k\ov w_{0} R_{1}} \cos\theta + {m\ov  R_{2}}
\sin\theta \bigg) y^0_1 
 +\ \bigg( -{k\ov w_{0} R_{1}}
 \sin\theta + {m\ov  R_{2}}
\cos\theta \bigg) y^0_2 \ , 
$$
The BPS condition that there are no oscillations along $y_2$ 
becomes 
$$
\a ^i_{(k,m)}=0 \ \ \ {\rm if} \ \ \ \  
R_{2} k \sin \theta \neq w_{0} R_{1} m \cos \theta \ ,
$$ 
i.e.  if $ k q_2 \not= w_0 m q_1$.  Thus the relevant states are
constructed    using  $\a _{(-k, -m_0)}^i $  with 
$$ 
m_0=   {  q_2 \ov  w_{0} q_1   } k \ .
$$
For  such  states,
$$
{ H_{\rm (A) } }=\sum _{k=1}^\infty \a^i_{(-k,-m_0)}\a^i_{(k,m_0)}\ , \ \ \ \ \
\ \ 
 \omega_{km_0}  = {k \ov q_1}\sqrt{  q_1^2 +  q_2^2 \tau _2^2 } \ ,
$$
and the constraints become 
$$
N_\s^+={q_1\ov  \sqrt{ q_1^2+ 
 q_2^2 \tau _2^2 } } \sum _{k=1}^\infty  \a^i_{(-k,-m_0)}\a^i_{(k,m_0)}
=l_1 w_{0} \ ,
$$
$$
N_\rho^+={q_2\ov  w_{0}\sqrt{ q_1^2+ 
 q_2^2 \tau _2^2 } } \sum _{k=1}^\infty  \a^i_{(-k,-m_0)}\a^i_{(k,m_0)}
=l_2 \ .
$$
The  membrane  BPS mass formula is then 
$$
M^2 = {l_1^2\ov R_{1}^2 }+ {l_2^2\ov R_{2}^2 }+
{w_0^2 R_{1}^2\ov {\a' }^2 } + {2 \ov \a' }H_{\rm (A)}\ ,
$$
with
$$
 H_{\rm (A)}= {1\ov q_1}  \sqrt{  q_1^2 +  q_2^2 \tau _2^2 }\ 
N_\s ^+ \ = \ 
w_0\sqrt{ l_1^2 +l_2^2 \tau _2^2} \ .
$$
Thus  
$$
M^2=\bigg( \sqrt { {l_1^2\ov R_{1}^2} +{l_2^2\ov R_{2}^2 } }
+ {w_0 R_{1}\ov \a'  } \bigg)^2 \ .
$$
Remarkably, this agrees  with the Schwarz string 
 mass formula (\ref{eq:35})
for BPS states with $N_L=0$.


\subsection{T-Duality in M-Theory}

In standard perturbative string theory, T-duality symmetry
is the assertion  that  conformal field theories corresponding to
two backgrounds related by a T-duality transformation are equivalent.
In particular, the one-loop partition function is the same 
for both systems, and 
there is a one-to-one correspondence between the spectra.
Related to these properties is the fact that the effective
action is T-dual invariant to all orders in the $\a' $ expansion.
In eleven dimensional supergravity, solutions with  three
 or more isometries
are also related by similar transformations \cite{berg}.
Whether this property is to hold beyond leading order in M-theory
remains to be proved.
In particular, it strongly relies on a complete
matching of the mass spectra of excitations of the dual backgrounds.
This is the case in string theory, where 
winding states are crucial in order
for T-duality to be an exact  symmetry to all orders in perturbation theory.

For string theory, T-duality is  a symmetry  of the world-sheet
action: given a $\s$-model background with some isometry, 
one can find the dual $\s$-model background by a standard procedure
based on gauging the isometry and introducing at the same time
Lagrange multipliers which set the gauge field to zero;
then the original space-time coordinates are removed by a gauge fixing,
and one obtains the $\s$-model action for the dual background;
up to some subtleties (periodicities, etc.)
the corresponding conformal field theories are guaranteed to be
equivalent \cite{rabin}.
A similar procedure in membrane theory does not work 
\cite{dulu,sezg,towss,ahar}.
In 2+1 dimensions the dual to scalars are vectors, and what one obtains
by this procedure
is {\it not}  a membrane theory on the dual background.
Given that T-duality symmetry seems to be inherent to two-dimensional
Lagrangians, it is natural to wonder why this should be expected
to be a symmetry of M-theory.
Consider the weak coupling limit $g^2\to 0$.
We can distinguish   quantum states states in the
spectrum with mass $\a'M^2=O(1)$ and those with $\a' M^2=O(1/g^2)$.
The former have a regular mass as $g^2\to 0$ and constitute the
perturbative string spectrum. The  exact T-duality of perturbative string theory implies that the spectrum in this sector is T-dual invariant. 
Less known is the sector of states with $\a' M^2=O(1/g^2)$.
Nevertheless,  we will explicitly see below that  the BPS subsector $\a' M^2\bigg|_{\rm BPS}=O(1/g^2)$
is also T-dual invariant
(the T-duality invariance of the BPS subsector is also ensured
by electric-magnetic duality of $N=4$  super Yang-Mills theory 
in 3+1 dimensions \cite{susski,taylor}).
This and  the invariance of the leading order effective action
are presently the only elements that support the idea of a T-dual M-theory.

For a direct test of T-duality in string theory, one may consider
the fundamental string background
in type IIA theory with a momentum boost along the string direction
and perform a T-duality transformation along the string direction, that is,
\beq
{\rm (a)}\equiv \ \upa +1_{NS} \ \mapa{T}\ 1_{NS}+\upa \ \equiv {\rm (b)}
\eeq
The full spectrum of oscillations of (a) and (b) backgrounds 
indeed coincide, after interchanging the winding charge
with the momentum and $R_1\to R_1'=\a'/R_1$.
The analogous test in M-theory can now be performed. In sect. 2 
we have seen that there are two backgrounds (A) and (B)
representing fundamental 2-branes
which are related by T-duality:
\beq
{\rm (A)}\equiv \ 2\nea \ \mapa{T_\bot T_{||} }\ 2\upa \ \equiv {\rm (B)}
\eeq
The T-dual backgrounds (A) and (B) are given by
\bea
{\rm (A):}\ \ \ \ \ \ 
 ds^2_{11}= && H_2^{-2/3}\big[
-dt^2+dy_1^2+dy_2^2+(\td H_2-1) (dt- dz_1)^2\big]
\nonu \\
&& +H_2^{1/3} (dy_3^2+dx_idx_i)\ ,
\nonu \\
 C_3=&& H_2^{-1} dt\wedge dy_1\wedge dy_2\ ,\ \ \ \ 
z_1=y_1 \cos\theta +y_2\sin\theta
\eea
\bea
{\rm (B):}\ \ \ \ \ \  
ds^2_{11}=&& \td H_2^{-2/3}\big[ -dt^2+dy_1^2+dz_2^2
+(H_2-1) (dt-dy_1)^2\big]
\nonu \\ 
&&+\td H_2^{1/3} (dz_3^2+dx_idx_i)
\nonu \\
C_3=&&\td H_2^{-1} dt\wedge dy_1\wedge dz_2\ ,\ \  
\eea
$$
z_2=y_2\cos\theta+y_3\sin\theta\ ,\ \  \ \ 
z_3=-y_2\sin\theta+y_3\cos\theta\ ,
$$
$$
H_2=1+{Q\ov r^5}\ ,\ \ \ \td H_2=1+{\td Q\ov r^5}\ ,\ \ \ 
\ Q= c_0 {w_0R_1\ov\a '}, \ \ \ \ \td Q= c_0{n\ov R_1}\sqrt{q_1^2+q_2^2\tau_2^2}
$$
Note that in the case $q_2=0$ the restriction of an 
extra isometry in $y_3$ can be removed (i.e. $y_3$ can be added to the $x_i$), since this coordinate does not play any special role. In this particular
case the backgrounds are equivalent under the exchange of momentum and winding charges ($Q\leftrightarrow \td Q$). 

From type IIA perspective, in going from one representation to another,
one has
$$
{\rm (A)}\ = \ \upa +0_{R}+ 1_{NS}\ \mapa{T_{y_1}} \ 1_{NS}+1_R+\upa\  
\mapa{T_{y_3}} \ 1_{NS} +2_{R}+ \upa \ = \ {\rm (B)}
$$
We have already calculated the spectrum of oscillations of the 2-brane
(A). We now calculate
the spectrum for the representation (B) \cite{trusso}.
An unboosted  $(q_1,q_2)$ string bound state is now represented
by a static 2-brane with one leg wrapped
around the coordinate $y_1$, and another wrapped around a $(q_1,q_2)$
cycle of the 2-torus
generated by $y_2, y_3$. That is:
$$
y_1(\s + 2\pi ,\rho )= y_1(\s ,\rho )+ 2\pi n R_1'\ ,\ \ $$
$$
y_2 (\s, \rho+2\pi )=y_2 (\s, \rho )+  2\pi q_1 R_2 \ , 
$$
$$
y_3 (\s, \rho+2\pi )=y_3 (\s, \rho )+  2\pi q_2 R_1 \ . \
$$
Adding the boost $w_0$ to the  $(q_1,q_2)$ string bound state
amounts to boosting along the coordinate $y_1 $ with momentum
$w_0/R_1'$. The target 3-torus coordinates can be expanded
as follows:
$$
y_1(\s, \rho )= n R_1' \s + \td y_1 (\s, \rho )\ ,
$$
$$
y_2(\s, \rho )= q_1 R_2 \rho + \td y_2 (\s, \rho )\ ,
$$
$$
y_3(\s, \rho )= q_2 R_1 \rho + \td y_3 (\s, \rho )\ ,
$$
$$
p_{y_1}=\int d\s d\rho \ P_{y_1} ={w_0\ov R_1'}\ ,\ \ \ \ \ 
p_{y_{2,3}}=\int d\s d\rho \ P_{y_{2,3}} = 0\ ,
$$
where $\td y_1, \ \td y_2,\ \td y_3$ are single-valued functions of $\s, \rho $.
Inserting this into the Hamiltonian,
\beq
H =  
2\pi ^2 \int d  \sigma d\rho \left [
 P _ a   ^ 2 + \ha 
{ T_3^2 }
( \{ X ^ a, X ^ b \} ) ^ 2\right ] \ ,
\eeq
and expanding all single-valued functions in $e^{ik\s +im\rho }$,
we now obtain 
\beq
\a'  H={ (R_1')^2 \ov 2\a '}(l_1^2 +l_2^2\tau_2^2 )+
{1\ov 2} \sum _{\n } \big[ P_{ \n }^a P^a_{-\n }
+\td \ww ^2 X^a_{ \n } X^a_{-\n} \big]+O\bigg({1\ov g}\bigg)
\ , \ \ \ \ 
\eeq
with
\beq
\td \omega _{km} = \sqrt{
k^2\bigg( q_1^2+q_2^2 {R_1^2\ov R_2^2} \bigg) +m^2 n^2{ R_1 ^{\prime 2}
\ov R_2^2} }\ .
\eeq
In the $g^2\to \infty $ limit, with  $\a ' $ and $R_1'/R_2$  fixed,
the (mass)$^2$ operator $M^2=2H -p_i^2$ then takes the form
\beq
M^2_{\rm (B)}=M_0^2 + {2\ov \a'} {  H_{\rm (B)}}\ ,
\label{eq:45}
\eeq
\beq
H_{\rm (B)}= \ha \sum _\n \big( \b^a_{-\n} \b^a_{\n} + \td \b^a_{-\n} \td \b^a_{\n}\big)\ ,
\eeq
\beq
[ \b _{ {(k,m)} }^a , \b^b_{(k',m')}]= \epsilon (k ) \td \ww \delta _{k+k'}\delta _{m+m'}\delta^{ab}\ , 
\eeq
with
\beq
N_\s^+ -N_\s^-= w_0n\ ,\ \ \ \ \ N_\rho^+= N_\rho ^-\ \ .
\eeq


For a BPS state, oscillations must be added along the momentum direction.
This
 implies that
the relevant states are made with the $\b_{(k,0)}^i, \ \td\b_{(k,0)}^i$
(or $\b_{(k,m)}^i=\td \b_{(k,m)}^i= 0$ if $m\neq 0$).
The BPS conditions (see sect. 4.1) also require that oscillations
are purely right moving, i.e. $\td\b_{(k,0)}^i=0$.
Then $N_\rho^+= N_\rho ^-= N_\s^-=0$, 
which indeed gives the minimum mass for given charges.
Thus $\td \omega _{km}^2 = k^2\big( q_1^2+q_2^2 \tau^2_2 \big)$, or
\beq
 H_{\rm (B)}=\sqrt{  q_1^2+q_2^2 \tau^2_2 } N_\s^+= w_0 
\sqrt{  l_1^2+l_2^2 \tau^2_2 }\ .
\label{eq:49}
\eeq
Substituting $R_1'=\alpha'/R_1 $ and eq.~(\ref{eq:49}) into
the mass operator (\ref{eq:45}), we obtain
\beq
M^2_{\rm (B)}= {l_1^2\ov R_1^2} + {l_2^2\ov R_2^2} + {w_0^2 R_1^2\ov 
\a ^{\prime 2}}
+{2\ov \a '}  w_0 
\sqrt{  l_1^2+l_2^2 \tau^2_2 }=\bigg( \sqrt{ {l_1^2\ov R_1^2}+{l_2^2\ov R_2^2} }+{w_0 R_1\ov \a '} \bigg)^2\ ,
\label{eq:500}
\eeq
which is in striking agreement with the   BPS spectrum of
the $(q_1,q_2)$ string bound states, and thus in agreement
with the BPS oscillation spectrum of the background (A),
as calculated in sect. 4.1.
For $l_2=0$, the matching with the  BPS sector of the spectrum (A)
is not a surprise:  in certain corners
of the modular parameters ($R_2\to 0$) we must recover exact 
T-duality of perturbative string theory. The BPS mass formula is exact and
should not receive additional corrections as we vary the radius.
But the spectrum (\ref{eq:500})  contains, in addition,
all quantum states with $l_2\in {\bf Z} \neq 0$ of masses  $M=O(1/R_2)$.
 The fact that  (A) and (B) BPS spectra match,
including these states $\a' M^2=O(1/g^2)$, 
constitutes a non-trivial test
of T-duality in M-theory.

In sect. 4.1, it was mentioned the  fact that the spectrum of $(q_1,q_2)$ strings is not
only reproduced for BPS states, but also for states containing both right and left moving oscillations along the momentum direction. 
Indeed, once the  $\b_{(k,m)}^i, \ \td \b_{(k,m)}^i$ with $m\neq 0$
are set to zero, the 2-brane mass spectrum becomes identical to the $(q_1,q_2)$
string spectrum,
\bea
&&M^2_{\rm (B)}\bigg|_{\rm along\  momentum}=
{l_1^2\ov R_1^2} + {l_2^2\ov R_2^2} + {w_0^2 R_1^2\ov \a  ^{\prime 2}}+4\pi T \sqrt{q_1^2+q_2^2\tau_2^2} (N_L+N_R)\ ,
\\
&& N_R-N_L=nw_0\ .
\eea
This is consistent with what one expects from the analysis of
the classical solutions: 
setting to zero the membrane excitations in the direction transverse
to the momentum is tantamount to the truncation of the spectrum
implied by the dimensional reduction, which indeed 
leads to the $1_{NS}+1_R+\upa $ string in type IIB theory.

\def\mbb{M_{\rm (B)}}
\def\maa{M_{\rm (A)}}

We now examine the problem of T-duality in non-BPS sectors.
The mass formulas for $\maa $ and $\mbb $
cannot be used to test T-duality in the non-BPS sectors, because
they apply in different corners of the torus modular parameters.
Although in both cases the relevant limit involves 
the strong coupling limit $g^2\to \infty $ with fixed $\a' $,
for (A) we have kept fixed $R_1/R_2$ (so that $R_1\to \infty $),
while for  (B)  we have fixed $R'_1/R_1=\a'/(R_1R_2)$ (so that
$R_1\to 0$).
Let us now take for (B) the limit at fixed $\tau_2=R_1/R_2$.
In this limit
\beq
\td \omega_{km}=\sqrt {k^2 \big( q_1^2+q_2^2 \tau_2^2 \big) +         {m^2n^2\over \tau_2^2 g^4} } \ \la \ k \sqrt { q_1^2+q_2^2 \tau_2^2 } \ .
\eeq
As a result, flat directions remain in the potential and one obtains
a continuum spectrum in this representation.
Instabilities are  produced by wave packets
constructed with the $X_{(0,m)}^a$, just as in sect. 3.
Along these directions the potential vanishes, and the wave packet can escape to infinity, leading to a continuum
spectrum of eigenvalues. In this (thin torus) $R_1'\to 0$ limit, the 
notion  of small 
oscillations around a stable configuration seems to break down for 
the 2-brane (B). In the dual description (A), where the membrane is wrapped
around a large area torus, there is nothing pathological and one
obtains an exact discrete spectrum.
If T-duality is a symmetry of M-theory, the true mass spectrum of
quantum states associated with background (B) at $R_1'\to 0$  
 must coincide  with that of representation (A); in particular, 
it must be discrete.

Similarly, 
in the strong coupling limit at fixed $R'_1/R_2$,
the mass spectrum of the 2-brane (A) becomes continuum, because
\beq
\ww =\sqrt { k^2 +  m^2  w_{0}^2 \tau_2^2 }\ \la\ k\ ,
\eeq
whereas in the same limit
one has an exact discrete spectrum for the 2-brane (B).

Thus, when the same limit is taken, the spectra of (A) and (B)
membranes do not match beyond the BPS sector.
Whether this should be attributed to a lack of T-duality in M-theory,
or to a breakdown of supermembrane theory (which is not renormalizable)
in this limit, or to something else, is unclear.
Nevertheless,  it may be fruitful to
explore the consequences of
including exact T-duality symmetry as part of an axiomatic definition of 
M-theory. 
In this simple approach, we may just 
demand $\mbb^2 =\maa^2$ and inquire what extra terms in the Hamiltonian
must be added.
For the sake of clarity, in what follows we consider the simpler
case $l_2=0$. In this case
we can write
\bea
 {\rm (A):}\ \ \ \ \ \ &&ds^2_{11}=-dt^2+dy_1^2+dy_2^2+(\td H_2-1) (dt- dy_1)^2+dx_idx_i\ ,\
\nonu \\
&& C_3=H_2^{-1} dt\wedge dy_1\wedge dy_2\ ,\ \ \ \ 
\label{eq:55}
\eea
and the background (B) is the same as eq. (\ref{eq:55}) with the exchange
$H_2 \leftrightarrow \td H_2$ (or $n \leftrightarrow w_0$, 
$R_1 \leftrightarrow \a'/R_1$).
Consider the limit $g^2\to \infty $, with $R_1/R_2$ fixed.
For the 2-brane (B), $w_0/R_1'$ represents the momentum $p_{y_1}$. 
Thus
$$
\ww =\sqrt { k^2 +  m^2  w_{0}^2 \tau_2^2 }=
\sqrt { k^2 +  m^2  {\a' p_{y_1}^2 \ov g^2} }\
$$
In the limit we are considering, both $g^2\to \infty $ and $p_{y_1}^2\to \infty $,
with the ratio  ${p_{y_1}^2 \ov g^2}$ fixed.
The Hamiltonian $\a' H$ must contain a term $\omega_{km}^2 X_{(k,m)}^2$.
The term $ k^2 X_{(k,m)}^2$ is already present in $\a' H$ from 
$$
\td \omega_{km}^2X_{(k,m)}^aX_{(-k,-m)}^a=\bigg(k^2+ {m^2n^2\ov \tau_2^2 g^4}\bigg)
X_{(k,m)}^aX_{(-k,-m)}^a\la  k^2 X_{(k,m)}^aX_{(-k,-m)}^a\ .
$$ 
Thus, the  term  $m^2  {\a' p_{y_1}^2 \ov g^2} X_{(k,m)}^2$ 
seems to originate from a Hamiltonian of the form
\beq
H=2\pi^2 \int d\s d\rho \ \big[ P^2_a+  T_3^2 R_2^2(\del_\s X^a)^2
+ T_3^2 n^2 R_1^{\prime 2}(\del_\rho X^a)^2+
{1\ov R_2^2} (\del_\rho X^a)^2 P_a^2 \big] +...
\label{eq:sinn}
\eeq
The last term is new and seems to be a concrete 
hint for a possible extension of supermembrane theory.

There is another good news about the existence of representation (B).
In the representation (A), we cannot
calculate the spectrum for a membrane with $w_0=0$, because
of membrane instabilities. 
Because of this, we were only able to check the correspondence with the BPS excitations
of the $(q_1,q_2)$ string in the sector $w_0\neq 0$, corresponding to
the string-string bound state with non-vanishing momentum $w_0/R_1'$.
In the representation (B),
it is possible to establish the correspondence between  string and membrane BPS spectra also in the sector
 $w_0=0$: now the membrane is stable
as long as $l_1\neq 0$ or $l_2\neq 0$, since in this case
the 2-brane winding is non-zero.
It is worth noting that the solution (A) with $w_0=0$
simply represents a gravitational wave;
if T-duality is to hold, the quantum states associated with this background 
in the limit $g^2\to\infty $,
$R_1\to 0$, can also be described  in terms of membrane excitations!
Although this circumvents the instability problem for this 
sector of the theory, 
the problem subsists for those states with $l_1=l_2=w_0=0$.

Let us summarize the results and give
 the spectrum in the different corners of
the modular parameters:

\noindent 1) $R_1, R_2\to \infty $, $T_3\to 0$
($g^2\to \infty $, with $\tau_2=e^{-\phi_0}=R_1/R_2$ and $\a' $ finite): 
\bea
&&\maa ^2=M_0^2 + {1\ov \a' }
\sum _\n \big( \a^a_{-\n} \a^a_{\n} + \td \a^a_{-\n} \td \a^a_{\n}\big)\ ,\
\ \ \ww =\sqrt { k^2 +  m^2  w_{0}^2 \tau_2^2 }\ , 
\nonu \\
&&N_\s^+ -N_\s^-= w_0n\ ,\ \ \ \ \ N_\rho^+= N_\rho ^-\ \ ,
\nonu \\
&&\mbb ^2={\rm continuum \ .}\ \
\nonu
\eea

\noindent 2) $R_2\to \infty $, $R_1\to 0$, $T_3\to 0$ 
($g^2\to \infty $, with $R_1'/R_2$ and $\a' $ finite)
\bea
&&\maa ^2={\rm continuum \ ,}\ \ \ 
\nonu \\
&&\mbb ^2= M^2_0+ {1\ov \a' }
\sum _\n \big( \b^a_{-\n} \b^a_{\n} + \td \b^a_{-\n} \td \b^a_{\n}\big)\ ,
\ \ \ \td \ww =\sqrt { k^2 +  m^2  n^2 { R_1 ^{\prime 2} \ov R_2^2} } \ ,
\nonu \\
&&N_\s^+ -N_\s^-= w_0n\ ,\ \ \ \ \ N_\rho^+= N_\rho ^-\ \ .
\eea

\noindent 3) $R_2\to 0 $,  $T_3\to \infty$ 
($g^2 \to 0$,  $\a' $ finite; this is the ten-dimensional limit)
\beq
M^2=M_0^2+  {2\ov \a' } (N_R +N_L)\ ,\ \ \ \ N_R-N_L=nw_0 \ .
\label{eq:57}
\eeq

There are  other possible ten-dimensional limits.
For example, 
in the limit  $R_1'\to 0$,  $T_3\to \infty $, with 
$\td \a ' \equiv (4\pi ^2 R_1' T_3)^{-1}$  fixed (note that 
$\td \a '=R_1R_2$),
 the relevant degrees of freedom of the system (B)
can be more adequately described by dimensionally reducing
on $y_1$, rather than on $y_2$.
In this process, we find
\bea
&&ds^2_{10A}=H_2^{1/2}\big[\td H_2^{-1} \big(-H_2^{-1}dt^2+
dy_2^2\big) + dy_3^2 + dx_idx_i\big]\ ,
\nonu \\
&&A=(H_2-1) H_2^{-1}dt\ ,\ \ \ \  B_{ty_2}=\td H_2^{-1}\ ,\ \ \ \ 
e^{2\phi }=\td H_2^{-1} H_2^{3/2}\ ,
\eea
which is a bound state of a D0-brane and a fundamental string.
The type IIA string coupling is $\td g^2=R_1'^2/\td \a ' $. 
A T-duality transformation along $y_2$ converts it into a R-R string with charge $w_0$
and a momentum boost $n/R_2'$, with $R_2' = \td\a '/R_2 $, i.e.
$$
0_R+1_{NS}\ \mapa{T_{y_2}}\ 1_R+\upa 
$$
According to eq. (\ref{eq:bosti}), the mass spectrum is then:

\noindent 4) $R_1'\to 0 $,  $T_3\to \infty$ 
($\td g^2 \to 0$,  $\td \a' =  (4\pi ^2 R_1' T_3)^{-1}$ finite)
\beq
M^2_{\rm RR}={n^2\ov R_2^{\prime 2}} + 4\pi^2 T^2_{(0,1)} {w_0^2 R_2^{\prime 2}} +
4\pi T_{(0,1)} (N_R+N_L)\ ,\ \ \ \ N_R-N_L=nw_0
\ ,
\eeq
with $T_{(0,1)}=(2\pi \td \a' )^{-1}{R_2\ov R_1'} $.
In terms of $\a '$, $R_1$, this becomes
\beq
M^2_{\rm RR}=  {n^2\ov R_1^2} + {w_0^2R_1^2\ov \a ^{\prime 2} }  
 +  {2\ov \a' } (N_R +N_L)\ ,
\label{eq:85}
\eeq
i.e. identical to eq. (\ref{eq:57}).

So far only two representations (A) and (B) have been  
considered.
Other duality transformations may be applied,
giving rise to new representations. An important question is
how many inequivalent membrane representations can be obtained
in this way. Let us again restrict our attention to the simplest case 
$l_2=0$.
As we have seen, in this case T-duality has a trivial
effect on the geometry; it just leads to an interchange of
 momentum and winding charges
associated with the direction of T-duality.
Thus, it will be sufficient to look at the transformation
of the radii: 
$$
(R_1,R_2) \ \ \mapa{T_{y_1}T_{y_3}}\ \ (R_1',R_2)\ \ ,\ \ \ \ 
R_1'={\a'\ov R_1}= {1\ov R_1R_2\bar T}\ ,
$$
$$
(R_1,R_2) \ \ \mapa{T_{y_2}T_{y_3}}\ \ (R_1,R_2')\ \ ,\ \ \ \ 
R_2'={\a'\ov R_2}= {1\ov R_1R_2\bar T}\ ,
$$
where $\bar T=4\pi^2 T_3$. The $T_{y_3}$ duality does not play any 
special role, and it will be implicit in what follows.
Then, the result we obtain is that the total number of
inequivalent membrane  representations is six, according
to the following scheme (see fig.~1):
\bea
1=(R_1,R_2) &&\mapa{T_{y_1}}\ \ 2= \bigg( {1\ov R_1R_2\bar T},R_2\bigg)
\ \ \mapa{T_{y_2}}\ \ 
3= \bigg({1\ov R_1R_2\bar T},R_1\bigg)\ \ \mapa{T_{y_1}}\ \
4=(R_2,R_1)
\nonu \\
 &&\mapa{T_{y_2}}\ \ 5=\bigg( R_2,{1\ov R_1R_2\bar T} \bigg)
\ \  \mapa{T_{y_1}}\ \  
6=\bigg( R_1, {1\ov R_1R_2\bar T} \bigg)\ \ \mapa{T_{y_2}}\ \
1=(R_1,R_2) \ ,
\nonu
\eea
S-duality connects $1 \leftrightarrow 4$, $2 \leftrightarrow 5$ and
$3 \leftrightarrow 6$.

\begin{figure}
{
\epsfxsize=9.0cm \epsfysize=7.7cm 
\epsfbox{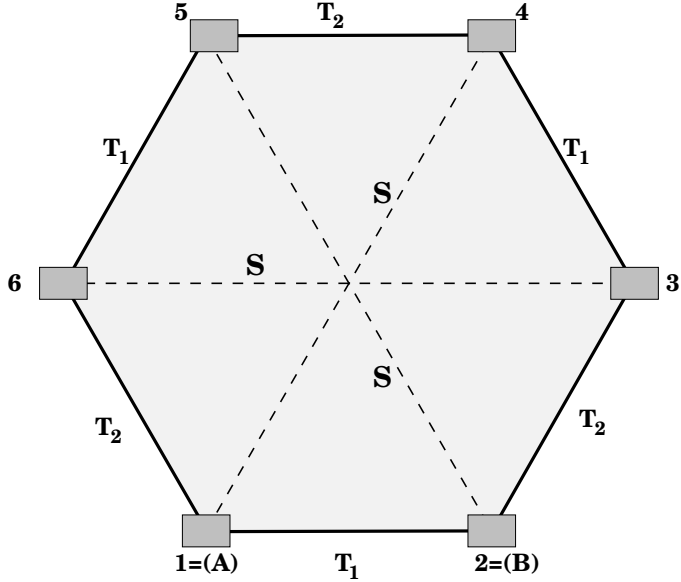} }
\caption{Alternate applications of T-duality 
in $y_1$ and $y_2$ directions lead to the same configuration
after six operations. The diametrically opposed points are 
connected by the simple $S$ duality that exchanges $y_1$ and $y_2$ 
directions.
}
\end{figure}

Assuming T-duality, 
the picture that seems to be emerging is shown in fig. 2.
Near the corners of the square 
at $T_3\to 0$ (inside the shaded surfaces), namely
$(R_1, R_2)\to (\infty , \infty) $,
and the other two corners related by $T$ and $S$ duality,
supermembrane theory is solvable and provides
an eleven-dimensional description that incorporates not only the
perturbative $(1,0)$ string states, but also the general $(q_1,q_2)$
oscillating quantum states (as well as the extra states of the membrane).
Indeed, we have seen that left and right moving oscillations along the momentum
direction of the 2-brane reproduce the full (BPS and non-BPS)
tower of states of string theory, with the only exception
of the  sector $w_0=l_1=l_2=0$, that we do not know
how to describe (this  
is related to the problem of  existence of a normalizable
ground state of the membrane Hamiltonian \cite{dewit,hoppe}).
The corners which are appropriate to the  membrane 
representations of fig.~1 are obtained by demanding that the associated 
target torus area is large. Using the values of the respective 
radii, we obtain the distribution as in  fig. 2.

\begin{figure}
{
\epsfxsize=11.0cm \epsfysize=10.3cm 
\epsfbox{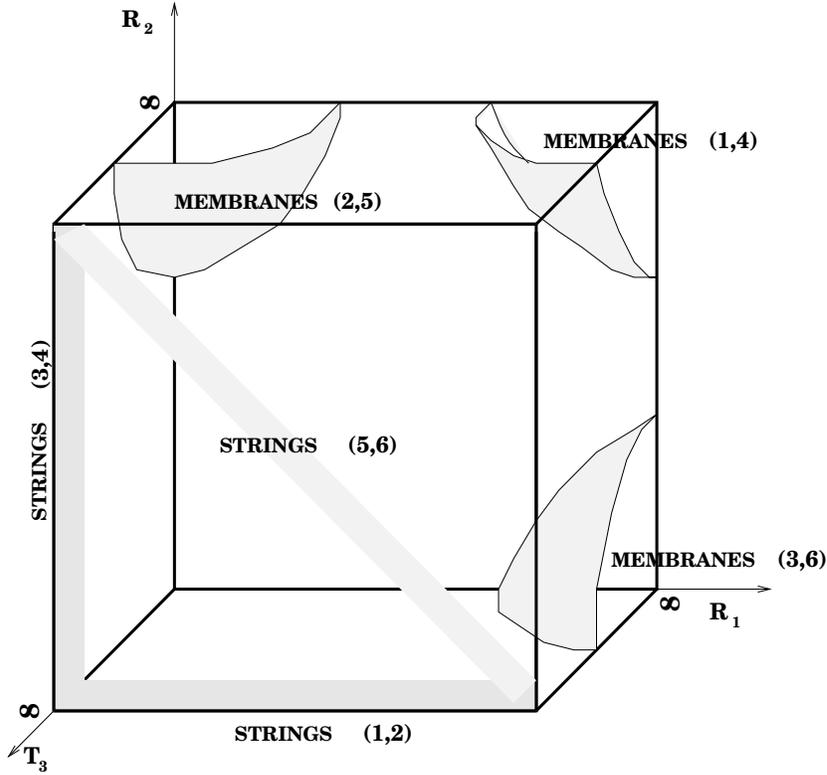} }
\caption{M-theory on ${\bf R^9}\times {\bf T}^2$.
The sides of the cube correspond to either 0 or $\infty $
values of the different parameters, 
as indicated in the figure. Only the parameter space for a rectangular ($\tau_1=0$) torus
is displayed.
The diagonal string zone corresponds to $R_1R_2=\a'$, arising, in
particular, in the limit that led to eq.~(\ref{eq:85}).
The numbers between parentheses in the string zones indicate
from what membrane such string description arises
upon reduction in $y_2$.
}
\end{figure}

At $T_3\to \infty$, various ten-dimensional limits can be taken, and we
recover the string spectrum, as we have just seen.
By using the full membrane Hamiltonian,
in sect. 4 it was  also  shown  
that in the limit that the torus area goes to zero 
at fixed $\tau_2^{-1} \ll 1$, the only 
membrane quantum states of perturbative mass that remain are those of the 
 superstring spectrum.
The different string descriptions of figure 2 arise as follows.
The horizontal edge is obtained by dimensionally reducing
${\rm (B)}=2$ or ${\rm (A)}=1$   along $y_2$ (as in eq.~(\ref{eq:57})); the  string zone in the vertical
edge follows from the reduction of membranes 3 or 4 along $y_2$
(as well as from (A) or 6 along $y_1$); 
the diagonal zone is obtained by reducing the membrane representations 
5 and 6 along $y_2$ (as well as from 3 or (B) along
along $y_1$, as in eq.~(\ref{eq:85})~).

5-branes are missing here. 
As in string theory --where there are  
soliton solutions representing higher-dimensional extended objects, 
but at weak coupling the $(1,0)$ string excitations are enough to define
a sensible perturbation theory--
in these corners of the moduli
space, it is not impossible
that the quantization of the 2-brane already provides the framework
for perturbative calculations in M-theory on ${\bf R}^9\times {\bf T}^2$.

\section {Conclusions}

We conclude by summarizing the main topics discussed here:

\noindent $\bullet $  Several classical solutions representing
(non-marginal) bound states of R-R and NS-NS $p$-branes admit a simple
$d=11$ description as a reparametrization of  configurations
of 5 and 2 branes. In eleven dimensions, the background
corresponding to a pure NS-NS configuration is no more
complicated that the U-dual version including R-R and NS-NS $p$-branes.

\noindent $\bullet $  The quantum states (as derived from
the light-cone Hamiltonian) of supermembranes wrapped
on  ${\bf R^9}\times {\bf T^2}$ that survive in the limit
of small torus area, at fixed $\tau_2^{-1}\ll 1$, are those of the 
type II superstring spectrum on ${\bf R^9}\times S^1$, in the sector
$w_0\neq 0$. For membranes on \rs , there remain extra quantum states.

\noindent $\bullet $  Correct BPS spectrum 
of the $(q_1,q_2)$ string bound state 
 can be derived from a
fundamental supermembrane, and in two inequivalent ways
(corresponding to M2-brane backgrounds related by T-duality).

\noindent $\bullet $ Full dynamics of wrapped $d=11$ 
2-branes can be understood in some corners of the  moduli space,
where   supermembrane theory becomes exactly solvable.

\noindent $\bullet $  If T-duality is an exact symmetry of M-theory,
boosted 2-branes with zero winding number
(which are unstable) may be described in terms of
a stable T-dual configuration.

\vskip .8cm

I am grateful to M.J. Duff, K. Stelle and A.A. Tseytlin for useful 
and stimulating discussions.
I would also like to thank the organizers of the APCTP Winter School
on String Dualities for their kind hospitality.

\end{document}